\documentstyle[prb,preprint,eqsecnum,aps]{revtex}
\tighten
\draft
\begin{document}
\title{Two - Dimensional Electron Liquid in a Weak  Magnetic Field}
\author{I. L. Aleiner and L. I. Glazman }
\address{
Theoretical Physics Institute, University of
Minnesota, 116 Church Str. SE, Minneapolis, MN 55455}
\date{Draft \today}
\maketitle
\begin{abstract}
We present an effective theory describing the low-energy properties of
an interacting 2D electron gas  at large  non-integer filling factors
$\nu\gg 1$. Assuming that the interaction is sufficiently weak, $r_s <
1$, we integrate out all the fast degrees of freedom, and derive the
effective Hamiltonian acting in the Fock space of the partially filled
Landau level only.  This  theory enables us to find two energy scales
controlling the electron dynamics at energies less than
$\hbar\omega_c$.  The first energy scale,
$(\hbar\omega_c/\nu)\ln\left(\nu r_s\right)$, appears in the one
electron spectral density as the width of a pseudogap. The second
scale, $r_s
\hbar\omega_c$, is  parametrically larger; it characterizes the
exchange-enhanced spin splitting and the  thermodynamic density of
states.
\end{abstract}
\narrowtext

\section{Introduction}
\label{sec:intro}
Since the discovery of the fractional quantum Hall effect\cite{Prange},
the properties of two-dimensional (2D) interacting electron systems in
a strong magnetic field have attracted persistent attention.
Historically, almost all the efforts were directed toward the study of
the strong magnetic field case when only the lowest Landau level is
occupied. Some attempts were undertaken \cite{Girvin86} to extend the
analysis to a larger filling factor $\nu >1$. However, they were all
limited to the case when the energy of the electron-electron
interaction is much smaller than the inter-Landau level spacing,
$\hbar\omega_c\gg e^2/(\kappa\ell)$, with  $\ell$ and
$\kappa$ being the magnetic length and the background dielectric
constant respectively. Under this condition, one can describe the
system in terms of one Landau level only; the effect of mixing of the
other Landau levels was either neglected or taken into account by
perturbation theory\cite{MacDonald84}.

In a weak magnetic field, the typical Coulomb energy exceeds the
cyclotron energy $\hbar\omega_c$. In this case, one can start from the
Fermi liquid theory in zero magnetic field.  The concept of Fermi liquid
usually enables one to neglect the interactions between quasiparticles
when determining  their energy spectrum in the vicinity of the Fermi
level.  This concept is based, first, on the screening of Coulomb
interaction, and, second, on the constraint on the phase space allowed
for a scattering event\cite{liquid}. Landau quantization of the
non-interacting quasiparticles in two dimensions creates a system of
discrete levels. The existence of energy gaps between the levels
affects adversely the screening, which involves the low-energy
excitations of the electron system. In addition, if the Fermi level
coincides with a Landau level, the system of non-interacting
quasiparticles becomes infinitely degenerate, and even a weak
interaction lifts this degeneracy. This may result in a non-trivial fine
structure of the partially filled Landau level (PFLL).

In this paper, we study the low-energy properties of the system with a
partially filled high Landau level\cite{MacDonald92}
 (the level index $N\gg 1$ equals to the integer part of $[\nu/2]$).  By
integrating out all the other degrees of freedom, we derive the
effective interaction between the electrons occupying this level.  This
procedure is valid for a weakly interacting  electron gas $r_s\lesssim
1$,  and at sufficiently large filling factors, $Nr_s \gg 1$. (Here $r_s
\equiv\sqrt{2} e^2/\kappa\hbar v_F$ is the conventional parameter
characterizing the interaction strength.)

The effective Hamiltonian enables  us to develop a microscopic theory
of the thermodynamic and tunnelling densities of states, and a
description of spin excitations.  In the tunnelling density of states, we
find  a pseudo-gap with a characteristic width
$\left(\hbar\omega_c/2N\right) \ln (Nr_s)$, which confirms the result
of the hydrodynamic approach\cite{Aleiner94}. The thermodynamic
density of states and spin excitations are characterized by the energy
scale $r_s\hbar\omega_c$.  Both  energies are smaller than
$\hbar\omega_c$.  It makes our theory complementary to the
conventional Fermi liquid picture\cite{Lee75} which is valid in the
energy range $E\gtrsim\hbar\omega_c$.

The paper is arranged as follows. Sec.~\ref{Peds} is a  qualitative
discussion of the energy scales relevant for the problem, and the
summary of our main results. Sec.~\ref{sec:Hamiltonian} is devoted to
the rigorous derivation of the effective Hamiltonian describing the
low-energy physics. In Secs.~\ref{sec:td} and \ref{sd} we apply the
effective Hamiltonian to study the thermodynamic and the tunnelling
density of states. The spin excitations are discussed in Sec.~\ref{sw}.

\section{Qualitative discussion and  results}
\label{Peds}
 Let us first consider an incompressible electron liquid with an integer
filling factor $\nu = 2N$. Because of the gap $\hbar\omega_c$ in the
spectrum, an external in-plane electric field ${\cal E}$ can not be
screened by the 2D electron system. Instead, it causes only a finite
polarization per unit area,
${\cal P}=\chi{\cal E}$. The polarizability of the incompressible 2D
electron liquid reduces the interaction $U(r)$ between two point
charges embedded into it,
\begin{equation}
U(r)=\int \frac{d^2{q}}{(2\pi)^2}\frac{2\pi e^2}{\kappa q}
\frac{1}{\varepsilon(q)}e^{i{\bf qr}},
\label{potentialq}
\end{equation}
where the two-dimensional dielectric function ${\varepsilon(q)}$ is
related to the polarizability $\chi$ by\cite{Keldysh}
\begin{equation}
\varepsilon(q) = 1 + \frac{2\pi q}{\kappa}\chi (q).
\label{chi}
\end{equation}

At small wave vectors, the matrix elements of the dipolar moment
between adjacent Landau levels, $d_{N,N-1}$, give the main
contribution to the polarizability,
$\chi\sim n_L|d_{N,N-1}|^2/\hbar\omega_c$; here $n_L=1/2\pi\ell^2$ is
the electron density on a Landau level. The characteristic spreading of
the electron wave function on a high Landau level is equal to the
cyclotron radius $R_c\equiv v_F/\omega_c$, and the estimate of the
dipolar moment is $d_{N,N-1}\sim eR_c$. Substituting $\chi$ into
Eq.~(\ref{chi}), we find:
\begin{equation}
\varepsilon(q) = 1 + \frac{R_c^2}{a_B}q,
\label{chilong}
\end{equation}
where $a_B=\hbar^2\kappa/me^2$ is the Bohr radius. Eq.~(\ref{chilong})
is valid only at small wave vectors, $qR_c\ll 1$. In the opposite limit,
$qR_c\gg 1$, a large number of Landau levels participate in
polarization of the 2D electron liquid. Therefore, the standard
Thomas-Fermi screening holds:
\begin{equation}
\varepsilon(q) = 1 + \frac{2}{a_Bq}.
\label{chishort}
\end{equation}
Formulas (\ref{chilong}) and (\ref{chishort}) match at $qR_c \sim 1$;
the corresponding value of the dielectric constant, $\varepsilon (q\sim
1/R_c)\sim R_c/a_B$, is large in the weak magnetic field limit, where
$R_c\gg a_B$.  The dielectric function for arbitrary $q$  is given by
Eq.~(\ref{simpleepsilon}), see Sec.~\ref{sec:Hamiltonian}.

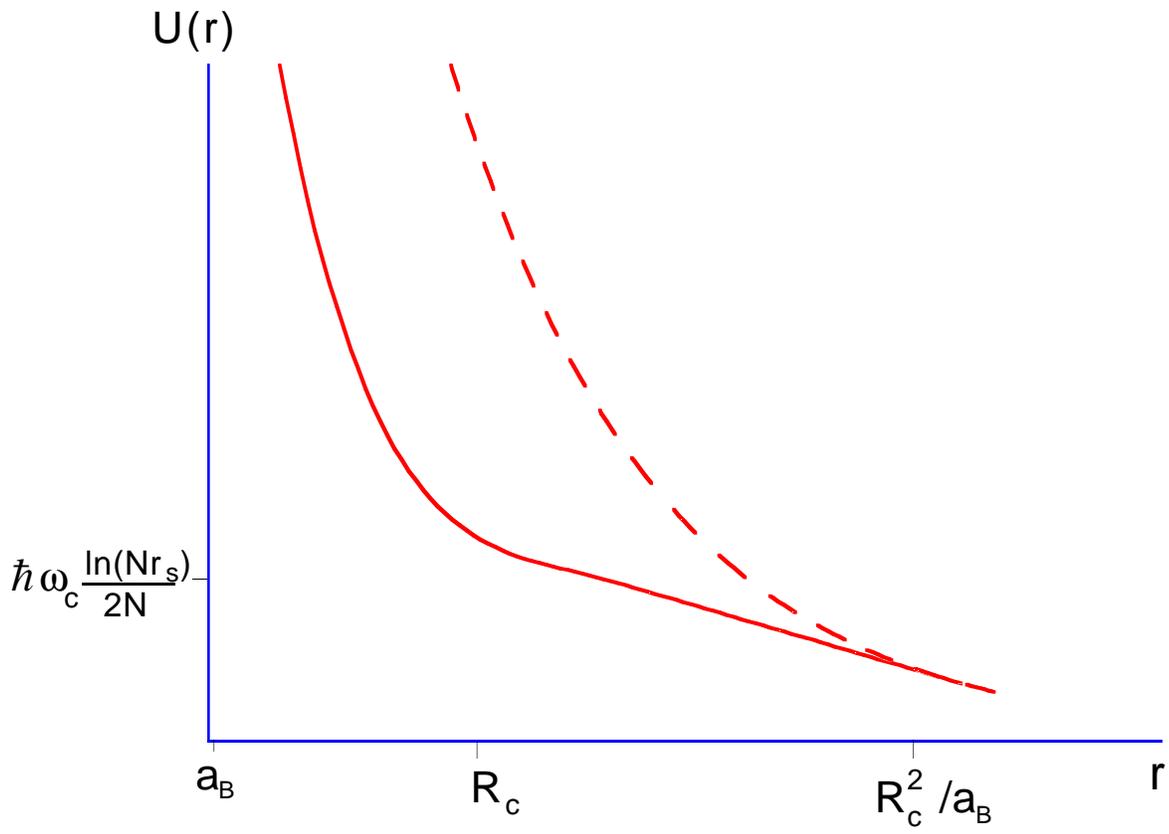
\begin{figure}
\caption
{The renormalized pair interaction $U(r)$ given by
Eq.~(\protect\ref{renpotential}) as the function of the inter-electron
distance (solid line). Dashed line is the bare Coulomb potential. The
asymptotic behavior of the renormalized potential is given by
Eqs.~(\protect\ref{potlong}) and (\protect\ref{potshort})}
\label{figpot}
\end{figure}

As it follows from Eq.~(\ref{chilong}), polarization is irrelevant only
for interactions on a very large length scale, $r\gg R_c^2/a_B$, where
$U(r)$ is given by the unscreened Coulomb interaction, see
Fig.~\ref{figpot}. At a smaller scale, $R_c\ll r\ll R_c^2/a_B$,
polarization is important, and Eqs. (\ref{potentialq}) and (\ref{chilong})
yield:
\begin{equation}
U(r) \simeq \frac{\hbar\omega_c}{2N}\ln\left(\frac{R_c^2}{a_B r}\right).
\label{potlong}
\end{equation}
A t $a_B \ll r\ll R_c$, according to Eq.~(\ref{chishort}), Thomas-Fermi
screening takes placer,  and the potential has the form
\begin{equation}
U(r) \simeq \frac{e^2a_B^2}{r^3}.
\label{potshort}
\end{equation}

The renormalized potential Eq.~(\ref{potentialq}) is significantly
smaller than $\hbar\omega_c$.  Therefore, unlike the bare Coulomb
potential, the renormalized interaction does not mix Landau levels. This
observation enables us to construct an effective theory of the low
energy properties of the electron system. Similar to the strong
magnetic field case, $\nu < 1$, the corresponding Hamiltonian is just
the energy of interaction between the electrons restricted to a single,
partially filled Landau level. The main difference is that  in our case
the interaction potential (\ref{potentialq}) is much smaller than bare
Coulomb  potential, and the PFLL has a high index $N\gg 1$. Although
still unsolvable, this low-energy theory is much simpler than the
original one, where all the Landau levels were relevant.

Armed with the effective theory, we are able to estimate the
characteristic energy scales which control the low-energy dynamics. In
order to make these estimates, we  employ the Hartree-Fock trial
function, based on the one-electron wave functions of the $N$-th Landau
level. (See Secs.~\ref{sec:td} and
\ref{sd} for more details.)   At small  electron concentration on the
PFLL,
$n_e^N\lesssim 1/R_c^2$,  the separation between electrons on this
level exceeds the size of the wave function, $R_c$. In this case the
system under consideration is equivalent to a classical crystal of point
charges interacting via potential (\ref{potentialq}). In the extremely
dilute limit,
$n_e^N\ll a_B^2/R_c^4$, the polarization is not important, and the usual
Coulomb repulsion between the electrons of the PFLL is not modified. In
this limit the electron system of the PFLL does not differ from a Wigner
crystal of electrons of the lowest Landau level in the strong magnetic
field regime. However, due to the polarization effect, there is a wide
region of electron densities in the weak field regime,
$a_B^2/R_c^4\lesssim n_e^N \lesssim 1/R_c^2$, where the Coulomb
repulsion in the crystal is replaced by the logarithmic potential
(\ref{potlong}). In this case, the static properties of the electron
crystal are equivalent to those of the vortex lattice in a thin
superconducting film\cite{super}.

We apply the developed picture to study of the energy dependence of the
tunnelling density of states. We find a suppression of the tunnelling
density of states (one-electron spectral density) at energies close to
the Fermi level
$E_F$, which is similar to the known
results\cite{Yang93WenHalperinWen2,Pikus,Kinaret} for a small filling
factor, $\nu\lesssim 1$. In the latter case, despite rather different
approaches, all the authors\cite{Yang93WenHalperinWen2,Pikus,Kinaret}
find a ``gap'' region of width $\Delta_g\simeq e^2n_e^{1/2}/\kappa$
(here
$n_e\equiv n_e^{N=0}$ is the total electron density at $\nu<1$). Such a
gap can be easily understood: $\Delta_g$ equals the Coulomb
interaction energy of an  extra electron with a ``frozen'' 2D electron
system. If there were no relaxation of the system at all, the density of
states would vanish at energies $|E-E_F| \lesssim\Delta_g$. The
relaxation processes smear out the threshold in the density of states;
however, the suppression is still strong at these energies, because the
characteristic relaxation time exceeds
$\hbar/\Delta_g$ in a strong magnetic field. For a weak magnetic field,
there are two stages in the evolution of the system after an electron
tunnelled in. On the short time scale, $t\sim\omega_c^{-1}$, the
polarization of the medium is formed, and the potential induced by the
additional electron acquires the form (\ref{potentialq}). At  larger time
scales, only the electrons of the PFLL can be re-distributed. The latter
processes are slow, which causes the suppression of the  tunnelling
density of states. The corresponding gap is determined by the
interaction of the tunneled electron with its nearest neighbors,
$\Delta_g
\simeq U(r \simeq1/\sqrt{n_e^N})$. For a broad range of the filling
factors of the PFLL,
$a_B^2/R_c^4\ll n_e^N\ll 1/R_c^2$, the width of the gap is:
\begin{equation}
\Delta_g \simeq
\frac{\hbar\omega_c}{4N}\ln\left(\frac{R_c^4n_e^N}{a_B^2}\right).
\label{gap1}
\end{equation} It is worth noticing that $\Delta_g$ depends on the
interaction strength and on the electron density on the PFLL only
logarithmically.

At larger concentrations, $n_e^N\gtrsim R_c^{-2}$, the single electron
wave functions are still orthogonal to each other, but the densities
corresponding to these wave functions have a significant spatial
overlap. Because of the strong repulsion between electrons at short
distances, one could naively expect that the interaction on these
distances gives the main contribution to the energy of the
many-electron state. However, for the fully spin polarized electron
system of the PFLL, the orbital part of the many-electron wave
function is antisymmetric, and vanishes whenever two electrons have
the same coordinates. This suppression of the wave function amplitude
compensates the large value of the interaction potential at $r\lesssim
a_B$. At larger distances,  $a_B\lesssim r\lesssim R_c$, the
interaction (\ref{potshort}) decays rapidly, and the corresponding
contribution to the Hartree-Fock energy turns out to be smaller than
the contribution of the long-range interaction (\ref{potlong}), see
Secs.~\ref{sec:td} and \ref{sd}. This results in the saturation of the
logarithmic growth of the gap $\Delta_g$ with the electron density: at
$n_e^N\gtrsim R_c^{-2}$ the gap in the tunnelling density of states is
\begin{equation}
\Delta_g \simeq  \frac{\hbar\omega_c}{2N}\ln\left(r_sN\right).
\label{gap2}
\end{equation}
This estimate was obtained previously  in the framework of the
hydrodynamic approach\cite{Aleiner94}.

We turn now to the discussion of the spin-flip excitations. The orbital
part of the many-electron wave function of such an excited state is no
longer antisymmetric. Therefore, the energy of such an excitation,
$\Delta_s$, is determined not only by  the bare Zeeman energy but also
by the extra interaction energy associated with the change in the
structure of the orbital wave function. Because the orbital part of the
wave function is not antisymmetric, the interaction on a short range
($r \lesssim a_B$) now does contribute to the energy of the state. This
corresponds to the well known exchange enhancement of the
$g$-factor\cite{Ando74}.  Following Ref.~\onlinecite{Kallin84} we
analyze the exactly solvable case of a PFLL which has a completely
filled spin-polarized sublevel. Flipping  the spin of a single electron
affects the exchange interaction of this electron with all the other
electrons of the {\em same} Landau level. All the lower Landau levels
carry equal number of electrons of both spin polarizations, and do not
contribute to the energy associated with the spin flip. Therefore, only
$1/2N$-th fraction of the total electron density participates in the
exchange enhancement, and the corresponding contribution to
$\Delta_s$ is approximately
$N$ times smaller than the exchange energy per electron at zero
magnetic field.  For a 2D electron gas  with Fermi wave vector $k_F$,
the latter energy is of the order $e^2 k_F/\kappa$, and the contribution
to $\Delta_s$ is of the order
$e^2 k_F/\kappa N$. The rigorous result obtained in Sec.~\ref{sw}
differs from this estimate only by a logarithmic factor.  For the
effective $g$-factor defined by the relation
\begin{equation}
g_{\rm eff}\equiv \frac{\Delta_s}{\hbar\omega_c},
\label{gdef}
\end{equation}
we obtain
\begin{equation}
g_{\rm eff}=g_0 + \frac{r_s}{\sqrt{2}\pi}
            \ln\left(\frac{2^{3/2}}{r_s}\right),
\label{geff}
\end{equation}
if the filling factor $\nu$ is odd. Here $g_0$ is determined by the
Zeeman splitting in the absence of the interaction, and it is usually
small (in
$GaAs$ the value of $g_0$ is only $0.029$). For the  even filling
factors, the value of $\Delta_s$ is determined solely by
$g_0$\cite{Ando74}.

We see that in the weak magnetic field, the energy scale of charge
excitations
$\hbar\omega_c/N$ is parametrically smaller than the energy scale for
spin excitations
$r_s\hbar\omega_c$. This is qualitatively different from the situation  at
low filling factor, $\nu\lesssim 1$, where both these excitations are
characterized by the same energy scale $e^2/\kappa\ell$.

\section{ Effective Hamiltonian}
\label{sec:Hamiltonian}
We start from the full Hamiltonian of the system
\begin{equation}
\hat{H}=\hat{H}_0 + \hat{H}_{int},
\label{initial_H}
\end{equation}
where the Hamiltonian for  non-interacting electrons,
 $\hat{H}_0$, is given by
\begin{equation}
\hat{H}_0=\sum_{n,k}\left[n\hbar\omega_c-\mu\right]
\psi_{n,k}^{\dagger}\psi_{n,k}.
\label{H0}
\end{equation}
Here $n$ and $k$ are the Landau level index and the guiding center
coordinate respectively (we adopt the Landau gauge). The fermionic
field operators in Eq.~(\ref{H0}) satisfy the standard anticommutation
relations
$\left\{\psi_{n,k},\psi_{m,q}\right\}=0,
\left\{\psi_{n,k}^{\dagger},\psi_{m,q}\right\}=\delta_{mn}\delta_{kq}$,
and
$\mu$ is the chemical potential (we include the energy
$\hbar\omega_c/2$ into
$\mu$). To simplify the notation, we omit the spin indices. We will also
neglect the Zeeman term $g_0\sigma\hbar\omega_c/2$ in all the
intermediate calculations. This is legitimate because
$g_0 \ll 1$ and therefore all the effects associated with the small
Zeeman energy can be included in the effective Hamiltonian after the
fast degrees of freedom are integrated out.

Electron-electron interaction is described by
\begin{equation}
\hat{H}_{int}=\frac{e^2}{2\kappa}
\int\!\!\int \frac{d^2\mbox{\boldmath $r_1$}d^2\mbox{\boldmath
$r_2$}}{\left|\mbox{\boldmath $r_1$}-\mbox{\boldmath $r_2$}\right|}
:\left(\hat{\rho}(\mbox{\boldmath $r_1$}) - n_e\right)
\left(\hat{\rho}(\mbox{\boldmath $r_2$}) - n_e\right):\quad ,
\label{HI}
\end{equation}
where $:\dots :$ stands for the normal ordering, $\hat{\rho}$ is the
electron density operator, $\hat{\rho}(\mbox{\boldmath
$r$})=\Psi^{\dagger}(\mbox{\boldmath $r$})
\Psi(\mbox{\boldmath $r$})$, and $n_e$ is the average electron
density. The electron annihilation operator
$\Psi$ is related to the operators $\psi_{n,k}$ by
$\Psi(\mbox{\boldmath
$r$})=
\sum_{n,k}\psi_{n,k}\varphi_{n,k}(\mbox{\boldmath $r$})$, where
$\varphi_{n,k}(\mbox{\boldmath $r$})$ are the single electron wave
functions in a magnetic field.

In order to develop the effective low-energy theory, we notice that the
expression for the partition function, ${\cal Z}=Tr \exp(-\beta
\hat{H})$, can be written as
\begin{equation}
{\cal Z}=Tr_N\left(Tr^{\prime}_N \exp(-\beta \hat{H})\right),
\label{z1}
\end{equation}
where $Tr_N$ is the trace over the Fock space of the PFLL ($N$ is the
index of the PFLL,
$\beta$ is the inverse temperature) and $Tr^{\prime}_N$ means trace
over all the other Landau levels. At low temperatures,
$\beta\hbar\omega_c\gg 1$, Eq.~(\ref{z1})  acquires the form
\begin{equation}
{\cal Z}=Tr_N\exp(-\beta \hat{H}_{eff}),
\label{z2}
\end{equation}
where the effective Hamiltonian $\hat{H}_{eff}$ is defined as
\begin{equation}
\hat{H}_{eff}=-\lim_{\beta_0\to\infty}\frac{1}
{\beta_0}\ln\hat{\Lambda}(\beta_0),
\quad
\hat{\Lambda}(\beta_0)=
Tr^{\prime}_N
\left\{
e^{-\beta_0\hat{H}}\right\}.
\label{Heff}
\end{equation}

Let us emphasize that, in general, the low frequency properties can be
described only by means of the effective action which includes a
retarded interaction. The description by an effective  Hamiltonian with
{\em instantaneous} interaction is possible only if the characteristic
time of the retardation is much smaller than $\hbar/E_c$, where $E_c$
is the maximal energy scale which is considered within the effective
theory. This condition is met for the problem under consideration.
Indeed, the summation in Eq.~(\ref{Heff}) involves inter-Landau level
transitions only. These transitions are associated with the energy gaps
not smaller than
$\hbar\omega_c$. Correspondingly, the characteristic time of the
retardation related to such a transition is $t\lesssim\omega_c^{-1}$.
On the other hand,  we saw at the Sec.~\ref{Peds} that
the largest characteristic energy $E_c=\Delta_s$ is still
parametrically smaller than
$\hbar\omega_c$, which enables us to neglect the retardation.

We now turn to the calculation of the effective Hamiltonian (\ref{Heff}).
The simplest way to proceed is to use the Hubbard-Stratonovich
transformation: first, we introduce auxiliary scalar field
$\phi(\tau,\mbox{\boldmath $r$})$ to decouple  quartic in $\Psi$
interaction part of the Hamiltonian; second, we perform the summation
over all the fermionic degrees of freedom not belonging to PFLL. After
that, we integrate out auxiliary fields by means of the saddle point
approximation and thus obtain the effective Hamiltonian
$H_{eff}$.

Below we implement this procedure. Performing the first step of the
Hubbard-Stratonovich transformation, we find for $\Lambda$ from
Eq.~(\ref{Heff})
\begin{eqnarray}
&&\displaystyle{
\hat{\Lambda}(\beta_0)\!=\!\!\int\!\!{\cal D}\phi
\exp\left({\frac{1}{2}\!\int\! d^3\mbox{\boldmath $\xi$}\!\!\int\!
 d^3\mbox{\boldmath $\xi^\prime$}
\phi(\mbox{\boldmath $\xi$})
\phi(\mbox{\boldmath $\xi^\prime$})
K(\mbox{\boldmath $\xi\! -\! \xi^\prime$})}\right)
\times }
\nonumber \\
&&\displaystyle{Tr^{\prime}_N
\left\{
e^{-\beta_0\hat{H}_0}T_{\tau}\exp\left({-\int d^3\mbox{\boldmath
$\xi$}\phi(\mbox{\boldmath $\xi$})
\left(\hat{\rho} (\mbox{\boldmath $\xi$})\!-\! n_e\right)}
\right)\right\},}
\label{decoupling}
\end{eqnarray}
where $T_{\tau}$ is the imaginary time ordering, and $3{\rm D}$ vector
$\mbox{\boldmath
$\xi$}$ represents time and space coordinates,
$\mbox{\boldmath $\xi$}=(\tau,\mbox{\boldmath $r$})$. The domain of
integration over
$\mbox{\boldmath $\xi$}$ is: $\tau \in [0,\beta_0],\ x \in [0, L_x],\ y
\in [0, L_y]$, where $L_{x}\times L_{y}$ is the geometrical size of the
2D system. We will omit the Planck constant in all the intermediate
calculations.

The measure of  the functional integral in Eq.~(\ref{decoupling}) is
determined by the condition
\begin{equation}
\int\!\!{\cal D}\phi\
\exp\left({\frac{1}{2}\!\int\! d^3\mbox{\boldmath $\xi_1$}\!\int\!
 d^3\mbox{\boldmath $\xi_2$}
\phi(\mbox{\boldmath $\xi_1$})
\phi(\mbox{\boldmath $\xi_2$})
K(\mbox{\boldmath $\xi_1 - \xi_2$})}\right)\!=\!1,
\label{normalization}
\end{equation}
and the function
$K$ is defined by the equation
\begin{equation}
\int d^3\mbox{\boldmath $\xi_3$}
K(\mbox{\boldmath $\xi_1\!-\! \xi_3$})
\frac{e^2\delta (\tau_3\!-\!\tau_2)}
{\kappa| \mbox{\boldmath $r_3\!-\! r_2$} |}
=\delta (\mbox{ \boldmath $\xi_1\!-\!\xi_2$}).
\label{K}
\end{equation}

The density operator is
\begin{equation}
\hat{\rho}(\mbox{\boldmath
$r$},\tau)=\bar{\Psi}(\mbox{\boldmath $r$},\tau +\epsilon)
\Psi(\mbox{\boldmath $r$},\tau), \quad \epsilon \to  +0,
\label{rho}
\end{equation}
where the fermionic operators in Matsubara representation are
$\Psi(\tau)=e^{\tau \hat{H}_0}\Psi e^{-\tau \hat{H}_0}$, and
$\bar{\Psi}(\tau)=\Psi^{\dagger}(-\tau)$. By introducing  the
infinitesimal positive time shift $\epsilon$ into definition (\ref{rho})
we preserve the normal ordering of the fermionic operators in the
interaction Hamiltonian, see Eq.~(\ref{HI}).

After the decoupling is performed, the fermionic part of the
Hamiltonian in Eq.~(\ref{decoupling}) becomes quadratic, which enables
us to make the second step of the Hubbard-Stratonovich
transformation, i.e. to carry out the trace
$Tr^{\prime}_N$ over the ``fast'' fermionic degrees of freedom. The
result depends solely on the variables belonging to the PFLL:
\begin{mathletters}
\widetext
\begin{eqnarray}
&&\displaystyle{\hat{\Lambda}(\beta_0)=e^{-\beta_0\hat{H}^N_0}
\!\!\int\!\!{\cal
D}\phi\  e^{-F_0\{\phi\}}T_{\tau}\left\{
e^{-\hat{F}_1\left\{\phi,\hat{\rho}_N \right\}}
e^{-\hat{F}_2\left\{\phi,\Psi_N, \bar{\Psi}_N \right\}}
\right\},} \label{ugly}\\
&&\displaystyle{F_0\{\phi\}=\!\int\!\!
d^3\mbox{\boldmath
$\xi$} f_{\phi}(\mbox{\boldmath $\xi, \xi$})-\frac{N}{\pi\ell^2}
\int\! d^3\mbox{\boldmath $\xi$}
\phi(\mbox{\boldmath $\xi$})} \nonumber\\
&&\displaystyle{\quad\quad -\frac{1}{2}\!\int\! d^3\mbox{\boldmath
$\xi_1$}\!\int\!
 d^3\mbox{\boldmath $\xi_2$}
\phi(\mbox{\boldmath $\xi_1$})
\phi(\mbox{\boldmath $\xi_2$})
\left(K(\mbox{\boldmath $\xi_1\!-\! \xi_2$})-4G_N(\mbox{\boldmath
$\xi_2$},\mbox{\boldmath $\xi_1$})
\tilde{G}_{\phi}(\mbox{\boldmath $\xi_1$},\mbox{\boldmath $\xi_2$})
\right),}
\label{F} \\
&&\displaystyle{\hat{F}_1\left\{\phi,\hat{\rho}_N \right\}=\int
d^3\mbox{\boldmath $\xi$}
\phi(\mbox{\boldmath $\xi$})\left(
\hat{\rho}_N (\mbox{\boldmath $\xi$})-n_e^N\right),} \label{F1}\\
&&\displaystyle{\hat{F}_2\left\{\phi,\Psi_N, \bar{\Psi}_N \right\}
=\!\int\! d^3\mbox{\boldmath $\xi_1$}
\int\! d^3\mbox{\boldmath $\xi_2$}
\left[
\bar{\Psi}_N(\mbox{\boldmath $\xi_1^+$})
\Psi_N(\mbox{\boldmath $\xi_2$})
- 2G_N(\mbox{\boldmath $\xi_2$},\mbox{\boldmath $\xi_1$})
\right]
\phi(\mbox{\boldmath $\xi_1$})
\tilde{G}_{\phi}(\mbox{\boldmath $\xi_1$},\mbox{\boldmath $\xi_2$})
\phi(\mbox{\boldmath $\xi_2$}),
} \label{F2}\\
&&\displaystyle{G_N(\mbox{\boldmath $\xi_1$},\mbox{\boldmath
$\xi_2$})=-P_N(\mbox{\boldmath
$r_1$},\mbox{\boldmath
$r_2$}) e^{(\tau_2-\tau_1)(N\omega_c-\mu)}\theta
(\tau_1-\tau_2-\epsilon)},
\label{GN}
\end{eqnarray}
\label{ugly1}
\narrowtext
\noindent
where the $3{\rm D}$ vector $\mbox{\boldmath $\xi$}^+$ is  $(\tau
+\epsilon,
\mbox{\boldmath $r$})$. In Eqs.~(\ref{ugly1})  $\Psi_N$ is
the projection of
the fermionic operator (in Matsubara representation) on the PFLL,
$\Psi_N(\mbox{\boldmath
$\xi$})=
e^{\tau(\mu -N\omega_c)}\sum_{k}\psi_{N,k}
\varphi_{N,k}(\mbox{\boldmath
$r$})$, operator $\bar{\Psi}_N(\tau)$ is defined as
$\bar{\Psi}_N(\tau)={\Psi}_N^{\dagger}(-\tau)$.
Projections of the density operator $\hat{\rho}$ and of the Hamiltonian
$\hat{H}_0$ on the PFLL are given by $\hat{\rho}_N(\mbox{\boldmath
$\xi$})=\bar{\Psi}_N(\mbox{\boldmath
$\xi$}^+){\Psi}_N(\mbox{\boldmath
$\xi$})$ and by
$\hat{H}^N_0=\left(N\omega_c-\mu\right)\int d^2\mbox{\boldmath
$r$}
\hat{\rho}_N(\mbox{\boldmath
$r$})$ respectively. In Eq.~(\ref{GN}),
$P_N(\mbox{\boldmath $r_2$},\mbox{\boldmath $r_1$})$ is the
projection operator onto the PFLL\cite{projector};
$n_e^N=n_e-N/\pi\ell^2$ is the average electron concentration on the
PFLL.   Operator
$\hat{F}_2$ in Eq.~(\ref{F2}) is defined in such a way
$T_\tau\hat{F}_2=\hat{F}_2$. This follows from the anticommutation
relation for the fermionic operators
$\left\{\Psi_N(\mbox{\boldmath $r_1$}),
{\Psi}_N^{\dagger}(\mbox{\boldmath
$r_2$})\right\}=P_N(\mbox{\boldmath $r_1$},
\mbox{\boldmath $r_2$})$.
\end{mathletters}

The Green's function $\tilde{G}_{\phi}(\mbox{\boldmath
$\xi_1$},\mbox{\boldmath
$\xi_2$})$ is the kernel of the integral operator:
\begin{equation}
\hat{G}_{\phi}=\hat{P}_{N}^{\perp}\left[-\frac{\partial}{\partial \tau}
+ \mu
+\frac{\omega_c}{2}-
\hat{P}_{N}^{\perp}\hat{\cal
H}_{\phi}\hat{P}_{N}^{\perp}\right]^{-1}\hat{P}_{N}^{\perp},
\label{G}
\end{equation}
Here $\hat{P}_{N}^{\perp}$ is the projection operator onto the space of
functions {\em orthogonal} to the states of the PFLL \cite{projector},
and
$\hat{\cal H}_{\phi}$ is the Hamiltonian of an electron in the magnetic
field and in the external potential $\phi$:
\begin{equation}
\hat{\cal H}_{\phi}=\frac{\left(-i\nabla +\frac{e}{c}\mbox{\boldmath
$A$}\right)^2}{2m}
+
\phi(\mbox{\boldmath
$r$},
\tau).
\label{Hphi}
\end{equation}
The  function  $\tilde{G}_{\phi}$ describes the evolution  in the
external field $\phi$ of  electron states constrained to the Landau
levels
$n\neq N$. This constraint is implemented by the  introduction of the
projection operators $\hat{P}_{N}^{\perp}$ into Eq.~(\ref{G}).   At
$\tau_1=\tau_2$  the function $\tilde{G}_{\phi}$ is defined as
\begin{equation}
\tilde{G}_{\phi}(\tau_1, \tau_1)  \to
\tilde{G}_{\phi}(\tau_1,\tau_1+\epsilon)
\label{equal_times}
\end{equation}
in accordance with Eq.~(\ref{rho}).

The first term in Eq.~(\ref{F}) results from the summation over the
fermionic states with $n\neq N$, and represents the  ``thermodynamic
potential'' of electrons in these states in the field $\phi$. The function
$f_{\phi}(\mbox{\boldmath $\xi_1, \xi_2$})$  is defined as the kernel
of the operator
\begin{equation}
\hat{f}_{\phi}=2\ln\hat{G}_{\phi},
\label{f}
\end{equation}
with  $\hat{G}_{\phi}$ from Eq.~(\ref{G}). Here, the factor $2$
comes from the  spin degeneracy.

In order to perform the integration over the auxiliary field in
Eq.~(\ref{ugly}), we employ the saddle point approximation.  It means
that one has to expand the functional $F_0\{\phi\}$, see Eq.~(\ref{F}),
and the operator $\hat{F}_2\left\{\phi,\Psi_N, \bar{\Psi}_N
\right\}$,  see Eq.~(\ref{F2}), up to the second order in $\phi$.

We start from the expansion of the operator
$\hat{F}_2\left\{\phi,\Psi_N,
\bar{\Psi}_N \right\}$. It contains explicitly a factor, which is bilinear
in
$\phi$, and therefore we can replace $\tilde{G}_{\phi}$  in
Eq.~(\ref{F2}) by the Green's function $\tilde{G}_{0}$ of an electron in
the absence of an external field, $\phi = 0$. Then, Eq.~(\ref{G}) enables
us to find the explicit form of $\tilde{G}_{0}$:
\begin{eqnarray}
&&\displaystyle{\tilde{G}_0(\mbox{\boldmath $r$}_1,\tau_1;
\mbox{\boldmath $r$}_2,\tau_2)
= \!\!\sum_{k,n\neq N}\varphi_{n,k}^{\ast}({\bf r}_2)
\varphi_{n,k}({\bf r}_1)
e^{(\tau_2-\tau_1) (n\omega_c - \mu)}}\nonumber\\
&&\displaystyle{\times\left[\theta(\tau_2-\tau_1)\theta(N-n) -
\theta(\tau_1-\tau_2)\theta(n-N)\right]}.
\label{G0}
\end{eqnarray}

When expanding the first term in
$F_0\{\phi\}$, we use definition (\ref{f}) and  the solution of
Eqs.~(\ref{G}), (\ref{Hphi}) up to the second order in $\phi$, and we
obtain:
\begin{eqnarray}
&\displaystyle{\int\! d^3\mbox{\boldmath $\xi$}
f_{\phi}(\mbox{\boldmath $\xi,
\xi$})\approx \beta_0 \Omega_0+\frac{N}{\pi\ell^2}
\int\! d^3\mbox{\boldmath $\xi$}
\phi(\mbox{\boldmath $\xi$}) +}& \nonumber \\
&\displaystyle{+\!\int\! d^3\mbox{\boldmath $\xi_1$}\!\int\!
 d^3\mbox{\boldmath $\xi_2$}
\phi(\mbox{\boldmath $\xi_1$})
\phi(\mbox{\boldmath $\xi_2$})
\tilde{G}_{0}(\mbox{\boldmath $\xi_2$},\mbox{\boldmath $\xi_1$})
\tilde{G}_{0}(\mbox{\boldmath $\xi_1$},\mbox{\boldmath $\xi_2$}),}&
\label{fe}
\end{eqnarray}
where
$\Omega_0=\left[(N-1)\hbar\omega_c/2-\mu\right]NL_xL_y/\pi\ell^2$
is the thermodynamic potential of the system of noninteracting
electrons on the completely filled Landau levels. The third term in  the
functional $F_0\{\phi\}$, see Eq.~(\ref{F}), is already bilinear in
$\phi$.

Substitution of Eq.~(\ref{fe}) into Eq.~(\ref{F}) yields
\begin{equation}
F_0\{\phi\}= \beta_0 \Omega_0 -
\frac{1}{2}\int\!\int\!d^3\mbox{\boldmath
$\xi_1$}d^3\mbox{\boldmath $\xi_2$}
\phi(\mbox{\boldmath $\xi_1$}) {\cal S}\left(\mbox{\boldmath
$\xi_1$}-\mbox{\boldmath $\xi_2$}
\right)\phi(\mbox{\boldmath $\xi_2$}),
\label{FE}
\end{equation}
where the second term describes  fluctuations of the field $\phi$ which
are renormalized due to the integrated out degrees of freedom. The
kernel ${\cal S}$ equals to
\begin{equation}
{\cal S}(\mbox{\boldmath $r$},\tau)=\int \frac{d\omega d^2{q}}
{(2\pi)^3}\frac{\kappa q}{2\pi e^2}
{\varepsilon(q,\omega)}e^{i{\bf qr}-i\omega\tau},
\label{renorm}
\end{equation}
where $\varepsilon(q,\omega)$ coincides with the dielectric function
calculated in the random phase approximation\cite{Mahan}:
\begin{equation}
\varepsilon(q,\omega)=1 -
\frac{2\pi e^2}{\kappa q}\Pi(q,\omega).
\label{fullepsilon}
\end{equation}
Here, the polarization operator $\Pi(q,\omega)$ in
Eq.~(\ref{fullepsilon}) is given by
\begin{eqnarray}
&&\displaystyle{\Pi(q,\omega)=
2\int d^2\mbox{\boldmath $r$}d\tau e^{-i{\bf qr}+i\omega\tau}
\left[\tilde{G}_{0}(\mbox{\boldmath
$r$}, \tau; 0, 0) +\right. }\nonumber\\
&&\displaystyle{\left.\stackrel{}{{G}_{N}}(\mbox{\boldmath
$r$}, \tau; 0, 0)\right]
\left[\tilde{G}_{0}(0, 0; \mbox{\boldmath $r$}, \tau)+G_N(0, 0;
\mbox{\boldmath $r$}, \tau]
\right],}\label{polarization}
\end{eqnarray}
where the functions $\tilde{G}_0$ and  $G_N$ are defined by
Eqs.~(\ref{G0}) and (\ref{GN}) respectively. The explicit form and the
asymptotic behavior of $\Pi(q,\omega)$ are presented in Appendix
\ref{ap1}.

Within the approximations made,  we can transform Eq.~(\ref{ugly}) to
the form of a Gaussian integral over the auxiliary field $\phi$.  The
result is
\begin{eqnarray}
&&\displaystyle{
\hat{\Lambda}(\beta_0)\!=\!
e^{-\beta_0(\Omega_0+\hat{H}^N_0)}
\!\!\int\!{\cal D}\phi\ T_{\tau}\!
\left[
e^{-\hat{F}_1\left\{\phi,\hat{\rho}_N \right\}-
\hat{F}_2^0\{\phi,\Psi_N, \bar{\Psi}_N\}}
\right]
}\nonumber \\
&&\displaystyle{\times\exp\left(\frac{1}{2}\!\int\!
 d^3\mbox{\boldmath $\xi_1$}\!\int\!
 d^3\mbox{\boldmath $\xi_2$}
\phi(\mbox{\boldmath $\xi_1$})
\phi(\mbox{\boldmath $\xi_2$})
{\cal S}(\mbox{\boldmath $\xi_1\! -\!\xi_2$})\right)},
\label{almostthere}
\end{eqnarray}
where $\hat{F}_2^0$ is obtained from operator $\hat{F}_2$,
see Eq.~(\ref{F2}), by replacing $\tilde{G}_\phi$ with $\tilde{G}_0$.

Before we proceed further, let us emphasize that the saddle point
approximation can be justified only if the characteristic value of the
fluctuations of the auxiliary field $\phi$ is small enough. The estimate
which will be presented later in this section  shows that the
expansion quadratic in
$\phi$  is parametrically valid for the weakly interacting electron
system at large filling factors.

Now, we are in the position to perform the actual integration in
Eq.~(\ref{almostthere}). Operator $\hat{F}_1$, see Eq.~(\ref{F1}),  is a
linear functional of
$\phi$, whereas operator $\hat{F}_2^0$, see Eq.~(\ref{F2}), is quadratic
in
$\phi$; because the fluctuations of field $\phi$ are small, the typical
value of
$\hat{F}_2$ is much smaller than that of $\hat{F}_1$. If we neglect the
term
$\hat{F}_2^0$ at all, the functional integration in
Eq.~(\ref{almostthere}) can be easily performed.  The calculation is
further simplified if we approximate
${\cal S}(\tau ) \simeq \delta (\tau) \int {\cal S} (\tau ) d\tau$. It is
valid approximation for the description of the low-energy dynamics of
electrons belonging to PFLL: $S$ is  rapidly decaying function of time:
${\cal S}(\tau) \propto \exp (-\omega_c|\tau|)$ at $\tau \gtrsim
\omega_c^{-1}$, as it follows from
Eqs.~(\ref{renorm})~-~(\ref{polarization}). The integration over field
$\phi$ then results in an exponential
$\exp\left(-\beta_0\hat{H}_{int}^{eff}\right)$, with the Hamiltonian of
density-density interaction,
\begin{eqnarray}
\hat{H}_{int}^{eff}=\frac{1}{2}\int\!\!\int {d^2\mbox{\boldmath
$r_1$}d^2\mbox{\boldmath $r_2$}}
U(\mbox{\boldmath $r_1$}-\mbox{\boldmath $r_2$})
\label{heint}\\
\times:\left(\hat{\rho}_N(\mbox{\boldmath $r_1$}) - n_e^N\right)
\left(\hat{\rho}_N(\mbox{\boldmath $r_2$}) - n_e^N\right):.\nonumber
\end{eqnarray}
The operator (\ref{heint}) acts within the Fock space of PFLL,
and the renormalized pair
interaction $U(r)$ is related to ${\cal S}(\mbox{\boldmath $\xi$})$ by
\begin{equation}
\int d^3\mbox{\boldmath $\xi_3$}
{\cal S}(\mbox{\boldmath $\xi_1\!-\! \xi_3$})
U( |\mbox{\boldmath $r_3\!-\! r_2$} |) =\delta (\mbox{ \boldmath
$r_1\!-\!r_2$}).
\label{us}
\end{equation}
Substitution of Eq.~(\ref{renorm}) into Eq.~(\ref{us}) yields:
\begin{equation}
U(r)=\int \frac{d^2{q}}{(2\pi)^2}\frac{2\pi e^2}{\kappa q}
\frac{1}{\varepsilon(q)}e^{i{\bf qr}}.
\label{renpotential}
\end{equation}
Here the static dielectric function ${\varepsilon(q)}\equiv
{\varepsilon(q,\omega=0)}$ describes the renormalization of the
 bare Coulomb
potential due to the integrated out degrees of freedom. This function
can be easily found with the help of Eqs.~(\ref{fullepsilon}) and
(\ref{stat}), and it has the form
\begin{equation}
\varepsilon(q)=1+\frac{2}{qa_B}\left[1-{\cal
J}_0^2\left(qR_c\right)\right]
\label{simpleepsilon}
\end{equation}
for the large filling factor $N \gg 1$ and for values of wave vector $q$
much smaller than Fermi wave vector, $k_F$. In
Eq.~(\ref{simpleepsilon}),
$a_B=\hbar^2\kappa/me^2$ is the  Bohr radius and
${\cal J}_0 (x)$ is the zeroth order Bessel function.  Asymptotic
behavior of the renormalized potential $U(r)$ was discussed in
Sec.~\ref{Peds},  see also Fig.~\ref{figpot}.

The term $\hat{F}_2^0$ from Eq.~(\ref{almostthere}) can be taken into
consideration by means of the perturbation theory using the small
parameters $r_s$ and $1/N$. We show in Appendix
\ref{ap0} that this operator generates   term of the effective
Hamiltonian which is linear in density. This linear term can be
trivially included into the free electron Hamiltonian $\hat{H}^0_N$ by a
shift of the chemical potential.

Finally,  substituting the resulting expression for $\Lambda$ in
Eq.~(\ref{Heff}) and restoring the Zeeman  energy term, we obtain the
expression for the effective Hamiltonian:
\begin{eqnarray}
&&\displaystyle{\hat{H}_{eff}=\Omega-
\mu^{\ast}\int d^2\mbox{\boldmath $r$}
\hat{\rho}_N(\mbox{\boldmath $r$}) +\hat{H}_{int}^{eff}+}\label{He}\\
&&\displaystyle{\frac{1}{2}g_0\hbar\omega_c\int d^2
\mbox{\boldmath
$r$}\left(\Psi^\dagger_{N,\downarrow}(\mbox{\boldmath
$r$}),\Psi_{N,\downarrow}(\mbox{\boldmath
$r$}) -\Psi^\dagger_{N,\uparrow}(\mbox{\boldmath
$r$}),\Psi_{N,\uparrow}(\mbox{\boldmath
$r$})\right),}
\nonumber
\end{eqnarray}
where  $\hat{H}_{int}^{eff}$ is defined by Eq.~(\ref{heint}), the explicit
expression for the shifted chemical potential $\mu^\ast$ is presented
in Appendix \ref{ap0}, see Eq.~(\ref{Mufinal}), and the
 thermodynamic potential $\Omega$ of the electrons on  the filled
Landau levels is   given by Eq.~(\ref{Omegafinal}). We will not need
concrete values of  $\mu^\ast$ and $\Omega$ in the further
calculations.

Hamiltonian (\ref{He}) is the main result of this section.  The physical
meaning of this Hamiltonian is that the low-frequency dynamics of the
system is described by the electrons belonging to the upper Landau
level. Interaction between these electrons is renormalized due to the
large polarizability of all the other Landau levels.

Let us now discuss the condition of the validity of  the approximations
we made. As we mentioned earlier, the saddle point approximation can
be justified only if the characteristic value of the fluctuations of the
auxiliary field $\phi$ is small enough. Below we  show that the
fluctuations  at all the spatial and time scales are parametrically
small for a weakly interacting electron  system at large filling factors.

The magnitude of the fluctuations  localized
within the spatial range $r$ and within the time interval $\tau$ can be
estimated from Eqs.~(\ref{almostthere}) and (\ref{renorm}) as
\begin{equation}
\left<\phi^2\right>_{\tau,r} \simeq \frac{\hbar}{\tau}
 \left(\frac{e^2}{\kappa r\varepsilon(r^{-1},\tau^{-1})}\right).
\label{estimate}
\end{equation}
Further estimates depend on the relation between the scale $r$ and the
cyclotron radius $R_c$.

If this spatial scale  is large, $r \gg R_c$, only the dipole transitions
between the nearest Landau levels are induced. The corresponding
off-diagonal  matrix elements of the Hamiltonian are of the order of
$(R_c/r)|\phi|$. The mixing of the Landau levels by the field  $\phi$ is
small if
\begin{equation}
\frac{\left<\phi^2\right>_{\tau, r}}{\left(\hbar\omega_c
+
\hbar/\tau\right)^{2} }
\left(\frac{R_c}{r}\right)^2\ll 1.
\label{shfl}
\end{equation}
Combining Eq.~(\ref{estimate}) with Eq.~(\ref{shfl}) and using the
results (\ref{fullepsilon}) and (\ref{hydrodynamic}) for the dielectric
function, we find the condition of small fluctuations
\begin{mathletters}
\label{cond}
\begin{equation}
r_s N \gg 1.
\label{cond1}
\end{equation}
Therefore, the long-range fluctuations are not ``dangerous'' in the weak
magnetic field regime.

Let us now analyze the short range fluctuations, $r \ll R_c$.  In this
case
 transitions between distant Landau levels are possible,  and the
requirement of the smallness of  fluctuations coincides with the
standard  one for the validity of the random phase approximation at
zero magnetic field,
\begin{equation}
r_s \ll 1.
\label{cond2}
\end{equation}
Thus, the saddle point approximation is valid for the weakly interacting
electrons in a weak magnetic field.
\end{mathletters}

In the following sections, we will apply the effective Hamiltonian
(\ref{He}) to describe various physical effects associated with the
PFLL.

\section{Ground state energy and  thermodynamic density of states}
\label{sec:td}

In this section, we evaluate the ground state energy  and
thermodynamic density of states $\partial n_e/\partial \mu$ as a
function of the  filling factor of the partially filled Landau level
$\Delta\nu = \nu -2N$.  We consider explicitly the case $\Delta\nu \leq
1$. System   at the filling factors $1< \Delta\nu \leq 2$ can be
analyzed with the help of  the electron-hole symmetry, and the   the
thermodynamic density of states for this case can be obtained from the
results for $\Delta\nu\leq 1$ by the replacement $\Delta\nu \to
2-\Delta\nu$.

We assume that the ground state is spin polarized at $\Delta\nu\leq1$,
and thus omit the irrelevant Zeeman term\cite{polarizedsublevel}. The
chemical potential
$\mu^\ast$ for the given electron concentration on the PFLL
$n_e^N=M/L_xL_y$ is found by differentiating the ground state energy of
Hamiltonian (\ref{He}) with respect to the number of electrons $M$:
\begin{equation}
\mu^\ast=\frac{\partial E_0(M)}{\partial M},\quad E_0(M)=
\frac{\left<0|\hat{H}_{int}^{eff}|0\right>_M}
{\left<0|0\right>_M},
\label{mu}
\end{equation}
where $\left.|0\right>_M$ is the wave function of the ground state of
the system with $M$ electrons on the PFLL.

Evidently, it is sufficient to consider only the filling factors
 $\Delta\nu\leq 1/2$. At larger filling factors, $1/2 < \Delta\nu \leq
1$, one can use the electron-hole transformation within the spin
sublevel,
$\Psi_N^h=\Psi_N^{\dagger}$,  and study the system of holes with the
filling factor $1-\Delta\nu$ described by the Hamiltonian
\begin{eqnarray}
&&\hat{H}_{eff}=\Omega-\mu^{\ast}\int d^2\mbox{\boldmath $r$}
\left(\frac{1}{2\pi\ell^2}-\hat{\rho}_N^h
(\mbox{\boldmath $r$})\right) +
\hat{H}_{int}^{eff},
\label{Hh} \\
 &&\quad\hat{H}_{int}^{eff}=-\Delta_{ex}\int d^2\mbox{\boldmath $r$}
\left(\frac{1}{4\pi\ell^2}-\hat{\rho}_N^h(\mbox{\boldmath $r$})\right)
 +\nonumber\\
&&\frac{1}{2}\!\int\!\!\!\int\!\!
{d^2\!\mbox{\boldmath
$r_1$}d^2\!\mbox{\boldmath
$r_2$}}U(\mbox{\boldmath $r_1$}\!-\!\mbox{\boldmath
$r_2$}) \!:\!\left[\hat{\rho}_N^h(\mbox{\boldmath $r_1$}) -
n_h^N\right]\!
\left[\hat{\rho}_N^h(\mbox{\boldmath $r_2$}) - n_h^N\right]\!:.
\nonumber
\end{eqnarray}
Here
$\hat{\rho}_N^h=\left(\Psi_N^h\right)^{\dagger}\Psi_N^h$ is the hole
density operator
$\hat{\rho}_N^h$ and
$n_h^N=1/(2\pi\ell^2)-n_e^N$ is the average density of holes at the
PFLL. The first term in  $\hat{H}_{int}^{eff}$ corresponds to the shift of
the chemical potential of the completely filled Landau level due to the
exchange interaction\cite{Ando74,Kallin84}:
\begin{equation}
\Delta_{ex}=2\pi\ell^2\int d^2r U(r)P_N(r,0)P_N(0,r).
\label{exch}
\end{equation}
Here $P_N(r_1,r_2)$ is the projector operator on the
PFLL\cite{projector}. The integration in Eq.~(\ref{exch}) can be
performed with the help of the explicit form (\ref{property}) for
$P_N(r_1,r_2)$. Under the conditions (\ref{cond}), the calculation yields
\begin{equation}
\Delta_{ex}=
\hbar\omega_c
\frac{r_s}{\sqrt{2}\pi}\ln \left(\frac{2^{3/2}}{r_s}\right).
\label{exch1}
\end{equation}

It follows from Eq.~(\ref{Hh}) that filling of one spin sublevel results in
the shift $-\Delta_{ex}$ of the chemical potential. This enables us to
find the average thermodynamic density of states which is defined by
the relation
\begin{equation}
\overline{\left(\frac{\partial n_e}{\partial\mu}\right)} =
\frac{1/2\pi\ell^2}{\mu(\Delta\nu=1-\delta)-\mu(\Delta\nu=\delta)},
\quad
\delta\to +0.
\label{avtd}
\end{equation}
With the help of Eq.~(\ref{exch1}), we obtain
\begin{equation}
\overline{\left(\frac{\partial n_e}{\partial\mu}\right)}=
- \frac{m}{\hbar^2}
\left[\sqrt{2}r_s
\ln \left(\frac{2^{3/2}}{r_s}\right)\right]^{-1}.
\label{thermdensity}
\end{equation}
It is worth noticing, that the thermodynamic density of states
(\ref{thermdensity}) does not depend on magnetic field at all.  We will
see below that for a broad range of filling factors of the PFLL $\Delta
\nu$, the actual value of the thermodynamic density of states is close
to the average one given by Eq.~(\ref{thermdensity}).

The true ground state of the Hamiltonian (\ref{He}) for non-integer
filling factors is not known. In order to estimate the ground state
energy, we use the Hartree-Fock trial function  analogous to that used
by  Maki and Zotos\cite{Maki83} for the lowest Landau level
\begin{eqnarray}
\left.|0\right>_M=\prod_{j=1}^M\hat{\psi}_{\bf
R_j}^\dagger\left.|0\right>_0,
\label{trialfunction} \\
\hat{\psi}_{\bf R}^\dagger=\int d^2\mbox{\boldmath $r$}
\Phi_{\bf R}(\mbox{\boldmath
$r$})\Psi_N^{\dagger}(\mbox{\boldmath $r$}),
\label{c_s_creation}
\end{eqnarray}
where $\Phi_{\bf R}(\mbox{\boldmath $r$})$ is the normalized
one-electron wave function of a coherent state\cite{Kivelson87} on the
PFLL with the guiding center localized about point ${\bf R}$. In the
Landau gauge,
$A_y=-Bx$, $A_x=0$, this function has the form
\begin{eqnarray}
&\Phi_{\bf R}(\mbox{\boldmath $r$})=\frac{1}
{\sqrt{2\pi\ell^2}}e^{-i[{\bf
r}\times {\bf R}]{\bf z}/2\ell^2+ixy/2\ell^2}
g\left({\mbox{\boldmath $r -
R$}}\right),& \nonumber\\ &g\left(\mbox{\boldmath
$r$}\right)=\frac{1}{\sqrt{N!}}\left(\frac{x-iy}{\sqrt{2}\ell}\right)^N
e^{-{\bf
r}^2/4\ell^2}.&
\label{coherentstate}
\end{eqnarray}

The overlap between two functions (\ref{coherentstate}) rapidly
decreases with the distance between their guiding centers:
\begin{equation}
\int d^2\mbox{\boldmath $r$}
\Phi_{\bf R_1}^\ast(\mbox{\boldmath $r$})
\Phi_{\bf R_2}(\mbox{\boldmath $r$})=e^{-i[{\bf
R_1}\times {\bf R_2}]{\bf z}/2\ell^2}e^{-({\bf R}_1-{\bf
R}_2)^2/4\ell^2}.
\label{overlap}
\end{equation}
Equation (\ref{overlap}) shows that the overlap is exponentially small
even when the distance between the guiding centers is smaller than
$R_c$ and the electrons are not separated in space.

For small filling factors $\Delta \nu < 1/2$, we can choose the guiding
centers
$\mbox{\boldmath $R$}_i$ separated by the distance much larger than
the magnetic length $\ell$ which enables us to neglect the
non-orthogonality of the coherent states. Then, the expression for the
energy $E_0(M)$ given by Eq.~(\ref{mu}) acquires a simple form
\begin{eqnarray}
&&\displaystyle{E_0(M)=
\frac{1}{2}\sum_{i\ne j}^{M}V_{HF}(|\mbox{\boldmath
$R$}_i -
\mbox{\boldmath $R$}_j|) -
\frac{Mn_e^N}{2}\!\int
d^2\mbox{\boldmath $r$}U(r),} \nonumber\\
&&\displaystyle{V_{HF}(R)=V_H(R)-V_F(R),}
\label{HF_energy}
\end{eqnarray}
where Hartree $V_H$ and Fock $V_F$ potentials are given by
\begin{eqnarray}
&&\displaystyle{V_H(R)\!=\!\!\!\int\!\!\!\int\!\!
{d^2\mbox{\boldmath
$r$}\, d^2\mbox{\boldmath
$r^\prime$}}\ U(\mbox{\boldmath $r$}\!-
\!\mbox{\boldmath $r^\prime$})
\left|\Phi_{0}(\mbox{\boldmath $r$})\right|^2
\left|\Phi_{\bf R}(\mbox{\boldmath $r^\prime$})\right|^2}
 \nonumber\\
&&\displaystyle{=
\int\frac{d^2{q}}{(2\pi)^2}\frac{2\pi e^2}{\kappa q}
\frac{1}{\varepsilon(q)}
\left[L_N\left(\frac{q^2\ell^2}{2}\right)\right]^2
e^{-q^2\ell^2+i{\bf qR}}}
\label{Hartree}
\end{eqnarray}
and
\begin{eqnarray}
&&\displaystyle{V_F(R)\!=\!\!\!\int\!\!\!\int\!\!
{d^2\mbox{\boldmath
$r$}\, d^2\mbox{\boldmath
$r^\prime$}}\, U(\mbox{\boldmath $r$}\!-
\!\!\mbox{\boldmath $r^\prime$})
\Phi_{0}^\ast(\mbox{\boldmath $r$})
\Phi_{\bf R}^\ast(\mbox{\boldmath $r^\prime$})
\Phi_{0}(\mbox{\boldmath $r^\prime$})
\Phi_{\bf R}(\mbox{\boldmath $r$})} \nonumber\\
&&\displaystyle{\!=\!\int\!\!
\frac{d^2{q}}{(2\pi)^2}\!\frac{2\pi e^2}{\kappa q}
\frac{1}{\varepsilon(q)}
\left[L_N\left(\frac{q^2\ell^2}{2}\right)\right]^2
e^{-q^2\ell^2-{\bf q}{\bf R}-{\bf R}^2/2\ell^2}},\nonumber\\
\label{Fock}
\end{eqnarray}
\narrowtext
\noindent
respectively. Here $L_N(x)$ is the Laguerre polynomial\cite{handbook}.
We used Eqs.~(\ref{renpotential}) and (\ref{coherentstate}), when
deriving Eqs.~(\ref{Hartree}) and (\ref{Fock}). Small corrections
appearing due to the non-orthogonality of the functions
(\ref{coherentstate}) were discussed in Ref.~\onlinecite{Maki83}.

Now, the energy given by Eq.~(\ref{HF_energy}) should be minimized
with respect to the positions of the guiding centers $\mbox{\boldmath
$R$}_j$. The best configuration corresponds to the guiding centers
arranged in a triangular lattice\cite{Maki83}. Below we present
analytical results for the energy of the variational ground state
(\ref{trialfunction}) in  different domains of filling factor $\Delta\nu$.

{\em  Dilute system,  $\Delta\nu \ll N^{-1}$} -- In this case, the lattice
constant $a =
\left(\sqrt{3}n_e^N/2\right)^{-1/2}$ is large, $a\gg R_c$, so that the
Fock potential (\ref{Fock}) is exponentially small and the Hartree  term
$V_H(R)$ coincides with the interaction potential $U(R)$. In this limit,
two situations may be distinguished.

For an extremely dilute system, $\Delta \nu \ll N^{-3}r_s^{-2}$, the
lattice constant is larger  than $R_c^2/a_B$, and therefore the Coulomb
potential is not renormalized. In this case, the ground state energy
coincides with that of the Wigner crystal on  a neutralizing background,
\begin{equation}
E_0(M)=-\alpha\frac{e^2}{\kappa\ell}\frac{M^{3/2}}{M_{\Phi}^{1/2}}
\label{ex_dilute_energy}
\end{equation}
where $\alpha=0.782\dots$ is a numerical constant \cite{Maradudin77},
and $M_{\Phi}=L_xL_y/2\pi\ell^2$ is the number of states on a Landau
level.  The use of Eq.~(\ref{mu}) immediately yields for the
thermodynamic density of states in this regime
\begin{equation}
\frac{\partial n_e}{\partial\mu}=-
\frac{0.542}{r_s}\frac{m}{\hbar^2}
\left(\frac{\Delta\nu}{N}\right)^{1/2}.
\label{ex_dilute_td}
\end{equation}
The negative thermodynamic density
of states for an analogous system  was considered earlier in
Ref.\onlinecite{Bello81}.

For a moderately dilute system $N^{-3}r_s^{-2}
\ll \Delta\nu\ll N^{-1}$, the typical distance between electrons on the
 PFLL is much smaller than $R_c^2/a_B$. At such distances the
potential of interaction is strongly renormalized by the screening and
it is given by Eq.~(\ref{potlong}). With the logarithmic accuracy the
energy of the system is given by
\begin{equation}
E_0(M)\simeq-M\frac{\hbar\omega_c}{8N}
\ln\left[N^3 r_s^2\frac{M}{M_\Phi}\right]
\label{dilute_energy}
\end{equation}
and, correspondingly, the thermodynamic density of states for
moderately dilute system takes the form:
\begin{equation}
\frac{\partial n_e}{\partial\mu}=-\frac{4}{\pi}\frac{m}{\hbar^2}
\left(\Delta\nu{N}\right).
\label{dilute_td}
\end{equation}

{\em  ``Dense'' limit,  $N^{-1} \ll \Delta\nu \ll 1/2$} -- In this case the
distance between the nearest guiding centers is smaller than  $R_c$.
The asymptotic behavior of the Hartree and Fock potentials depends on
the relation between the magnetic length $\ell$ and the Bohr radius
$a_B$. We restrict ourselves to the case  $\ell \gg a_B$, or
\begin{equation}
N r_s^2 \gg 1,
\label{cond3}
\end{equation}
which enables us to  simplify Eqs.~(\ref{Hartree}) and (\ref{Fock}),
\begin{eqnarray}
&&\displaystyle{V_H(R) = \frac{e^2a_B}{2\pi R_c R} +
\frac{3e^2
a_B}{4\pi^2R_c^2}\ln\left(\frac{R_c}{R}\right) + \frac{ e^2
a_B}{R_c^2}\ln\left(\frac{R_c}{a_B}\right)}, \nonumber\\
&&\displaystyle{V_F(R) = \frac{e^2a_B}{2\pi R_cR
\left(1+Ra_B/2\ell^2\right)}}
\label{asymptotics}
\end{eqnarray}
for the range of  distances $\ell \ll R \ll R_c$.

The resulting potential $V_{HF}(R)=V_H(R)-V_F(R)$ is a smooth
function of
$R$ at $R\ll \ell^2/a_B$. This  enables us to find the analytical
expression for the ground state energy (\ref{HF_energy}) in the region
$N^{-1}r_s^{-2}\ll \Delta\nu \ll 1/2$. First, we can approximate
$\sum_{j\neq 0}V_{HF}({\bf R}_j)\approx n_e^N\int d^2RV_{HF}(R)
-V_{HF}(a)$, as  the potential varies only logarithmically over  the
lattice cell.  Second, we notice that the spatial average of the Hartree
potential is exactly equal to the average of the bare potential $U(r)$,
and these two terms cancel each other in Eq.~(\ref{HF_energy}) for
$E_0(M)$. Finally, the integral  of the Fock potential is proportional to
the exchange shift,
$\Delta_{ex}$, see Eq.~(\ref{exch1}). The resulting  energy of the
system is
\begin{eqnarray}
&\displaystyle{E_0(M)=-
\frac{M^2}{2M_\Phi}{\Delta_{ex}-}}& \label{dense_energy} \\
&\displaystyle{\, \ M\frac{\hbar\omega_c}{4N}
\left[\frac{3}{8\pi^2}\ln\left(\frac{M}{M_\Phi}N\right)+
\ln\left(Nr_s\right)+\frac{1}{\sqrt{2}\pi r_s}\right],}&
\nonumber
\end{eqnarray}
which yields the thermodynamic density of states of the form
\begin{eqnarray}
&\displaystyle{\frac{\partial n_e}{\partial\mu}=-
\frac{m}{\hbar^2}
\left[\sqrt{2}r_s
\ln
\left(
\frac{2^{3/2}}{r_s}
\right)
\right]^{-1}\times}& \nonumber \\
&\displaystyle{\left[1-\frac{3\sqrt{2}}
{32\pi r_s\ln \left(r_s^{-1}\right)}
\left(\Delta\nu N\right)^{-1}\right].}&
\label{dtd}
\end{eqnarray}
The leading term in Eq.~(\ref{dtd}) coincides
with Eq.~(\ref{thermdensity}).
\narrowtext

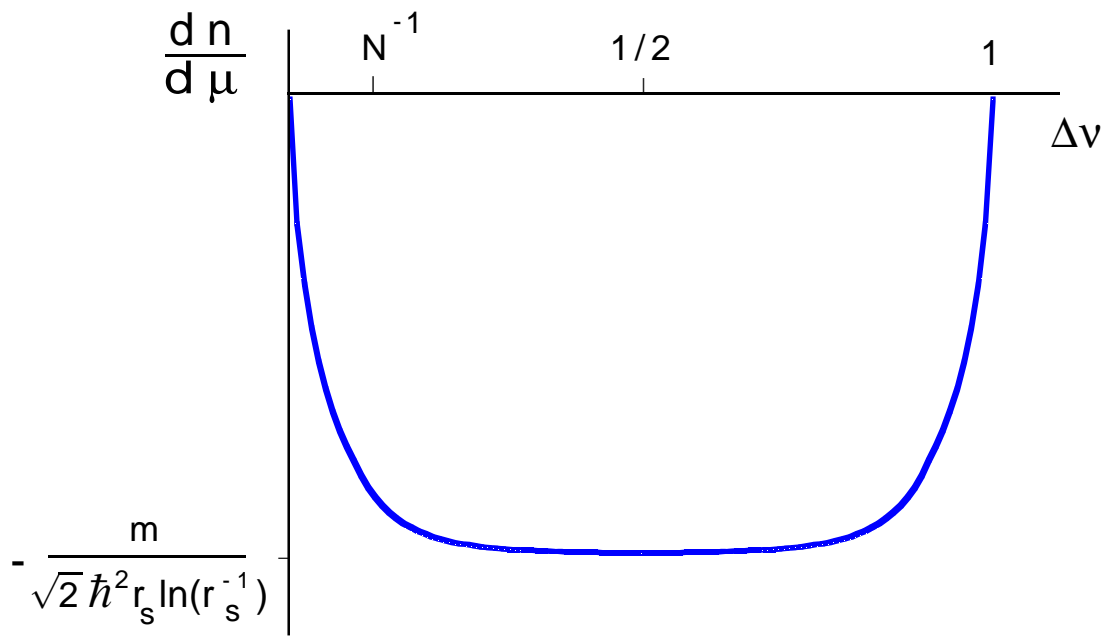
\begin{figure}
\caption {The thermodynamic density of states as a function of the
filling factor of the partially filled Landau level in the domain
$0 \leq \nu \leq 1$. The asymptotic behavior of the
thermodynamic density of states is given by
Eqs.~(\protect\ref{ex_dilute_td}), (\protect\ref{dilute_td}), and
(\protect\ref{dtd})}.
\label{fig2}
\end{figure}

The filling factors $\Delta\nu > 1/2$ can be considered  similarly by
using the electron-hole symmetry,  see Eq.~(\ref{Hh}), and the trial
function of type (\ref{trialfunction}). The overall dependence of the
thermodynamic density of states on the filling factor is shown in
Fig.~\ref{fig2}.

\section{Spectral density}
\label{sd}
The value of the one electron spectral density $A(\epsilon )$ can be
measured in the tunnelling experiments\cite{tunnel,weaktunnel}. For the
electron states  on the PFLL,
$A(\epsilon)$ is defined by relations:
\begin{eqnarray}
&&\displaystyle{A(\epsilon) = A^+(\epsilon) + A^-(\epsilon)},
\label{sddef} \\
&&\displaystyle{A^+(\epsilon)=\frac{1}{M_\Phi}\sum_{k,\tilde{m}}
\left|\left<\tilde{m}\left|\psi_{k,N}^\dagger\right|0\right>\right|^2
\delta (\tilde{E}_{\tilde{m}} -\tilde{E}_{0} - \epsilon), } \nonumber
\\
&&\displaystyle{A^-(\epsilon)=\frac{1}{M_\Phi}\sum_{k,\tilde{m}}
\left|\left<\tilde{m}\left|\psi_{k,N}\right|0\right>\right|^2
\delta (\tilde{E}_{0}-\tilde{E}_{\tilde{m}} - \epsilon).} \nonumber
\end{eqnarray}
Here, $\left|\tilde{m}\right>$ is an eigenstate of the {\em full}
Hamiltonian (\ref{initial_H}), $\tilde{E}_{\tilde{m}}$ is the
corresponding eigenvalue and
$\tilde{m}=0$ stands for the ground state of the system. Functions
$A^+(\epsilon)$ and
$A^-(\epsilon)$ describe the introduction of an extra electron or an
extra hole onto the PFLL respectively.

We are interested in the  spectral density at energies $|\epsilon | \ll
\hbar\omega_c$. All the behavior of the system at such low energies
can be described by the effective Hamiltonian (\ref{He}). Following the
method of Sec.~\ref{sec:Hamiltonian}, we can express the spectral
density (\ref{sddef})  in terms of the eigenstates of the {\em effective}
Hamiltonian  as
\begin{eqnarray}
&&\displaystyle{A^+(\epsilon)=\frac{Z}{M_\Phi}\sum_{k,m}
\left|\left<m\left|\psi_{k,N}^\dagger\right|0\right>\right|^2
\delta (E_{m} -E_{0} - \epsilon), }\nonumber
\\
&&\displaystyle{A^-(\epsilon)=\frac{Z}{M_\Phi}\sum_{k,m}
\left|\left<m\left|\psi_{k,N}\right|0\right>\right|^2
\delta (E_{0}-E_{m} - \epsilon),}
\label{SDEFF}
\end{eqnarray}
where $m$ denotes the eigenstates of the  effective Hamiltonian
(\ref{He}) and $E_{m}$ is the corresponding eigenvalue, $m=0$ is the
ground state. The $Z$-factor in Eq.~(\ref{SDEFF}) describes the overlap
between the low-energy  eigenstates of the full Hamiltonian and the
eigenstates of the effective Hamiltonian. The value of $Z$-factor
coincides with the quasiparticle weight for a $2{\rm D}$ degenerate
plasma in the absence  of the magnetic field,
\[
Z = 1-  \frac{3 r_s}{\pi \sqrt{2}}.
\]
Now we use Eq.~(\ref{SDEFF}) to evaluate the spectral density. Our
analysis is based on the trial function (\ref{trialfunction}). Even though
we are not able to find the details of the energy dependence of the
spectral density, this approach provides a reliable estimate for the
energy scales involved.

Because the trial function (\ref{trialfunction}) is constructed from the
coherent states (\ref{coherentstate}), it is convenient to rewrite
Eq.~(\ref{SDEFF}) in terms of the operators $\psi_{\bf R}, \psi_{\bf
R}^\dagger$, see Eq.~ (\ref{c_s_creation}), creating or annihilating an
electron in the coherent state:
\begin{mathletters}
\label{sdc}
\begin{eqnarray}
&&\displaystyle{A^\pm(\epsilon)=\frac{Z}{2\pi}\int dt e^{-i\epsilon t}
\int\frac{d^2\mbox{\boldmath
$R$}}{L_xL_y} {\cal G}^\pm ({\bf R}, t)}, \label{sdcs} \\
&&\displaystyle{{\cal G}^+ ({\bf R}, t) =
\left<0\left|e^{-it\hat{H}_{eff}}\psi_{\bf R}
e^{it\hat{H}_{eff}}\psi_{\bf R}^\dagger
\right|0\right>, }\label{sdcs1}
\\
&&\displaystyle{{\cal G}^- ({\bf R}, t) =
\left<0\left|\psi_{\bf R}^\dagger
e^{-it\hat{H}_{eff}}\psi_{\bf R}
 e^{it\hat{H}_{eff}}\right|0\right>.}\label{sdcs2}
\end{eqnarray}
\end{mathletters}
Here we used the  representation of
the projection operator in terms of the coherent
states (\ref{coherentstate}):
\[
P_N(\mbox{\boldmath $r_1, r_2$}) =\int
\frac{d^2\mbox{\boldmath $R$}}{2\pi\ell^2}
\Phi_{\bf R}^\ast(\mbox{\boldmath $r_2$})
\Phi_{\bf R}(\mbox{\boldmath $r_1$}).
\]

Let us concentrate below on the calculation
of the spectral density for the hole
excitations
$A^-(\epsilon)$.
(The  calculation procedure for  $A^+(\epsilon)$ is similar and will be
briefly outlined later.)

At small time $t$, expression (\ref{sdcs2}) can be approximated by
the formula
\begin{equation}
 {\cal G}^- ({\bf R}, t) \approx {\cal G}^- ({\bf R}, 0)
\exp\left( i t E^-({\bf R}) -
\frac{1}{2} \left[ tD ({\bf R})\right]^2\right)
\label{sdcs_exp}
\end{equation}
with  parameters
\begin{eqnarray}
&& \displaystyle{
E^- ({\bf R}) =\frac{\left<0\left|\psi_{\bf R}^\dagger,
\left[\psi_{\bf R}, {\hat{H}_{eff}}\right]
\right|0\right>}{\left<0\left|
\psi_{\bf R}^\dagger\psi_{\bf R}
\right|0\right>}};\label{E}
\\
&&
\displaystyle{ \left[D({\bf R})\right]^2
 =\frac{\left<0\left|\psi_{\bf R}^\dagger
\left[{\hat{H}_{eff}},\left[{\hat{H}_{eff}},\psi_{\bf R} \right]\right]
\right|0\right>}{\left<0\left|\psi_{\bf R}^\dagger
\psi_{\bf R}
\right|0\right>} - \left[E^- ({\bf R})\right]^2 .}
\nonumber
\end{eqnarray}
Substitution of Eq.~(\ref{sdcs_exp}) into Eq.~(\ref{sdcs})
gives the result
\begin{equation}
A^-(\epsilon)\!=\!\frac{Z}{\sqrt{2\pi }}
\!\int\!\!\frac{d^2\mbox{\boldmath
$R$}}{L_xL_y}\frac{{\cal G}^- ({\bf R}, 0)}{D ({\bf R})}\!
\exp\left\{-
\frac{\left(E^-({\bf R})\! -
 \!\epsilon\right)^2}{2 \left[D({\bf R})\right]^2}
\!\right\}
\label{sdfinal}
\end{equation}
for the spectral density.

Now we use Maki-Zotos trial
state (\ref{trialfunction}) to find the functions in the
integrand   in  Eq.~(\ref{sdfinal}).
At a small filling  of the PFLL, $\Delta\nu \ll 1$,
we obtain with the help of Eq.~(\ref{overlap})
\begin{equation}
{\cal G}^-({\bf R}, 0) =
\sum_{{\bf R}_i}\! e^{-({\bf R }-{\bf R }_i)^2/2\ell^2},
\label{densities}
\end{equation}
which describes the ``density of the centers
of orbits'' in the Wigner crystal.
Energy $E^-({\bf R})$ defined in Eq.~(\ref{E})  turns
 out to be independent
on the position
${\bf R}$ because all the electrons
 in the Wigner crystal have the same energy:
\begin{equation}
 E^- =
\sum_{{\bf R}_i \ne 0}V_{HF}(|{\bf R}_i|)
- n_e^N \!\int
d^2\mbox{\boldmath $r$}U(r) - \mu^\ast.
\label{energies}
\end{equation}
Here the Hartree - Fock energy  $V_{HF}$
is defined by Eq.~(\ref{HF_energy}), and the
second term describes the neutralizing background.
The parameter $D({\bf R})$ is also independent of $R$ and it is given by
\begin{equation}
D^2 = \ell^2 \sum_{{\bf R}_i \ne 0}
\left[\mbox{\boldmath $\nabla$} V_{HF}(|{\bf R}_i|)\right]^2.
\label{width}
\end{equation}
Because none of the parameters in the exponent in Eq.~(\ref{sdfinal})
 depend on ${\bf
R}$, the integration can be easily performed, and it yields:
\begin{equation}
A^-(\epsilon)\!=\Delta\nu\frac{Z}{\sqrt{2\pi }D}
\exp\left\{-
\frac{\left(E^-\! - \!\epsilon\right)^2}{2 D^2}
\!\right\}.
\label{sdfinal1}
\end{equation}

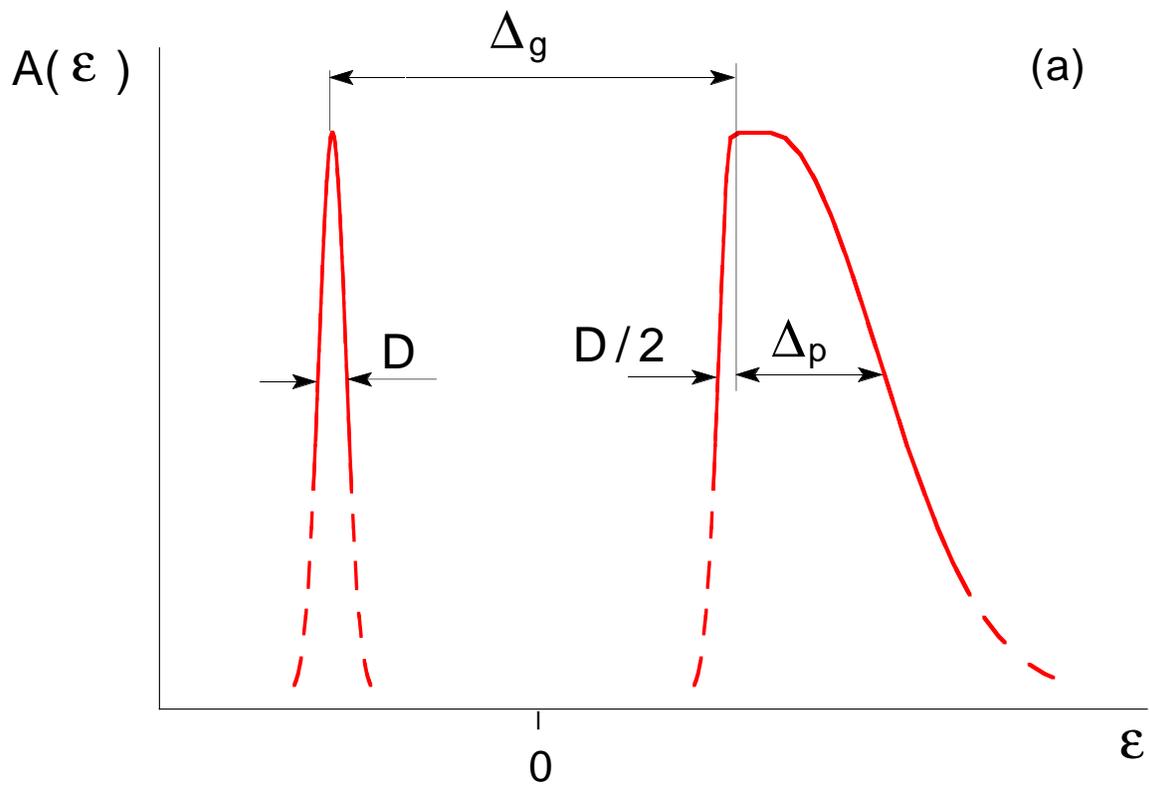
\begin{figure}
\caption { (a): The spectral density as a function of energy. (b): The
dependence of the characteristic energy scales  on filling factor of
the partially filled Landau level
$\Delta\nu$. The asymptotic behavior of the width of the pseudogap,
$\Delta_g$ is given by Eq.~(\protect\ref{gapas}). The width of the peak
in the one-electron spectral density
$\Delta_p$ is determined by the variation of the interstitial energy and
it is smaller than $\hbar\omega_c/2N$ at all filling factors.}
\label{figds}
\end{figure}

Let us now discuss the physical meaning of the energies $E^-$, $D$.
After
an electron is removed from the PFLL, the system acquires
extra energy equal to
$-E^-(\bf R)$ due to the interaction between the  hole  and all
 the  electrons of the PFLL.
If all the other electrons of the PFLL were ``frozen'',  this
state would give a  contribution
$\propto \delta (\epsilon - E^- ({\bf R}))$ to the spectral density.
(Classical model of the frozen electrons  was
 used in Ref. \onlinecite{Pikus} for the
electrons on the lowest Landau level.)
However, the state formed right after the
tunnelling is not an eigenfunction.
The decay of this initial state leads to the finite
quantum width ${D}$ of the spectral density, see Fig.~\ref{figds} a.

To find the region of  applicability of expression (\ref{sdfinal1})
we notice that the expansion in $t$ we used in Eq.~(\ref{sdcs_exp})
 is valid for $|t| \ll |E^-/D^2|$.  The characteristic time determining the
value of  the integral in Eq.~(\ref{sdcs}) can be estimated as $t \simeq
|\epsilon - E^-|/D^2$.  Combining these two conditions, we find that
Eq.~(\ref{sdfinal1}) is applicable for $|\epsilon - E^-| \ll |E^-|$. At the
boundary of  applicability, the spectral density (\ref{sdfinal1}) is
proportional to
$\exp\left[-(E^-/D)^2\right]$, and therefore  Eq.~(\ref{sdfinal1})
describes
the main contribution to the spectral density only if $|E^-| \gg D$.
Substituting the explicit expression for Hartree-Fock potential
(\ref{HF_energy}) into Eqs.~(\ref{energies}) and (\ref{width}),
we see that
the ratio
$D/|E^-|$ is maximal at $\Delta\nu\simeq1/2$, where its value is
$D/|E^-|\simeq 0.1(\ln N)^{-1/2} \ll 1$. Furthermore, we believe that the
use of the Hartree-Fock function sets the upper limit for $D$. This is
because trial function (\ref{trialfunction}) does not take into account
the correlations in the motion of centers of orbits, and thus
overestimates  zero-point fluctuations of the electron density. The use
of a more sophisticated\cite{Girvin84} trial function, or of a
phenomenological model that has the correct spectrum of  excitations
of the Wigner crystal\cite{Kinaret} would restrict the summation in
Eq.~ (\ref{width}) to the nearest neighbors only, and further reduce the
value of
$D$.

The spectral density of holes has a sharp peak, because a hole
corresponds to a vacancy in the Wigner crystal.  In contrast, the
tunnelling electron may create an interstitial at an arbitrary position
in the elementary cell.  This gives rise to a finite width of the spectral
density $A^+(\epsilon)$, even  for the ``frozen'', $D=0$, crystal. The
energy band for an interstitial is given by
\begin{equation}
E^+({\bf R}) =
\sum_{{\bf R}_i}\!\left[\frac{V_{HF}(|{\bf R}_i-{\bf R}|)}
{1\!-\!e^{-({\bf R }-{\bf R }_i)^2/2\ell^2}}\right]\!- \!n_e^N \!\int
\!d^2\mbox{\boldmath $r$}U(r) - \!\mu^\ast.
\label{E+}
\end{equation}
The leading term in the quantum width $D$ for electrons does not
depend on
${\bf R}$, and it is still given by Eq.~(\ref{width}).

To calculate the spectral density   $A^+(\epsilon)$, one can use
Eq.~(\ref{sdfinal}) with
$E^-({\bf R})$ and ${\cal G}^- ({\bf R}, 0)$ replaced by $E^+({\bf R})$ and
$1-{\cal G}^- ({\bf R}, 0)$ respectively. The width of the electron
spectral density is determined by the energy band (\ref{E+}).  The
abrupt edge of the spectral  density at
$\epsilon = {\rm min}_{\bf R}\left\{E^+({\bf R})\right\}$ is smeared
only by a small quantum width
$D$, see Fig.~\ref{figds} a.

The minimal energy necessary to create an interstitial, ${\rm
min}\left\{E^+({\bf R})\right\}$, is of the order of energy  $|E^-|$ needed
to create a vacancy. Therefore the difference
$\Delta_g={\rm min}\left\{E^+({\bf R})\right\} - E^-$
exceeds significantly $D$ and thus determines the pseudogap in the
spectral density (\ref{sddef}).  Using Eqs.~(\ref{energies}), (\ref{E+}),
and the asymptotic behavior of the Hartree-Fock potential $V_{HF}(R)$,
see Sec.~\ref{sec:td}, we obtain the width of the pseudogap for
different filling factors:
\begin{equation}
\Delta_g =  \left\{
\begin{array}{ll}
\displaystyle{\simeq \frac{e^2}{\kappa \ell}}\sqrt{\Delta\nu }, \quad
&\Delta\nu \ll N^{-3}r_s^{-2}\\
\displaystyle{\frac{\hbar\omega_c}{4N}
\ln\left[N^3 r_s^2\Delta\nu\right]}, \quad & N^{-3}r_s^{-2}\ll  \Delta\nu \ll
N^{-1}\\
\displaystyle{\frac{\hbar\omega_c}{2N}
\ln\left(Nr_s\right)},  \quad &  N^{-1} \ll \Delta\nu \ll 1/2,
\end{array} \right..
\label{gapas}
\end{equation}
\narrowtext
\noindent
The leading term in the width of the gap agrees with the result of
Ref.\onlinecite{Aleiner94} for the broad range of filling factors
$\Delta\nu \gtrsim 1/N$. The overall dependence of the characteristic
energy scales on the filling factor $\Delta\nu$ is shown in
Fig.~\ref{figds} b.

It is important to notice that in the dominant region of filling factors,
$\Delta\nu > 1/N$,  the width of the gap $\Delta_g$ is determined by the
renormalized interaction (\ref{potlong}) at distances $r\simeq R_c$,
which are large compared to the characteristic distance between the
electrons of PFLL. The gap is insensitive to the short range correlations
in the wave function of the ground state. Therefore, we anticipate that
the value of $\Delta_g$ we found is robust: it is not an artifact of the
Maki-Zotos trial function, but it should be the same (with logarithmic
accuracy) for any state homogeneous on the macroscopic scale
$r\gtrsim R_c$.

\section{Spin excitations}
\label{sw}
In this section we analyze the simplest excitations of the PFLL at
$\Delta\nu=1$. This problem is similar to the one studied in
Ref.~\onlinecite{Kallin84}. The only difference is that the interaction
potential is renormalized due to the polarizability of the other Landau
levels. Our goal is to study how this renormalization affects the spin
excitations.

The energy of the fully spin polarized ground
state\cite{polarizedsublevel}
$\left.|M=M_\Phi,S_z=M_\Phi/2\right>$  at $\Delta\nu =1$ is
\begin{equation}
E_p = \Omega - M_\Phi \left(\mu^\ast+\frac{\Delta_{ex} +
 g_0 \hbar\omega_c}{2}\right),
\label{Ep}
\end{equation}
where  $\Delta_{ex}$ is defined by Eq.~(\ref{exch1}).
The eigenstates of the Hamiltonian with an extra electron or hole
\[
\left.|e\right>=
\psi_{N,k,\downarrow}^\dagger\left|M_\Phi,\frac{M_\Phi}{2}\right>,
\quad
\left.|h\right>=\psi_{N,k,\uparrow}\left|M_\Phi,\frac{M_\Phi}{2}\right>
\]
have the energies
\begin{eqnarray}
&E_\downarrow =
 E_p - \mu^\ast + \frac{1}{2} g_0 \hbar\omega_c,&
\nonumber\\
&E_\uparrow =
 E_p + \mu^\ast + \Delta_{ex} + \frac{1}{2} g_0 \hbar\omega_c&
\label{eh}
\end{eqnarray}
respectively.  Equations~(\ref{eh}) enable us to relate
the width of the spin gap
$\Delta_s= E_\downarrow + E_\uparrow - 2E_p$
 to the value of $\Delta_{ex}$, i.e. $\Delta_s=\Delta_{ex}+g_0
\hbar\omega_c$. Substituting $\Delta_s$  in Eq.~(\ref{gdef})
 and using Eq.~(\ref{exch1}), we obtain
the result (\ref{geff}) for the effective $g$-factor.

It is worth noticing that the effective $g$-factor
 is independent on $N$ in the limit of a
weak magnetic field due to the Thomas-Fermi
 screening\cite{gfactor}.  (Without
screening, this factor would logarithmically diverge with $N\to\infty$.)

The exchange enhancement is maximal if the filling factor is odd. When
the both spin sublevels are either empty or completely filled ($\nu$ is
even), the spin splitting is determined by the bare $g$ -
factor\cite{Ando74}.

Let us turn to the consideration of  neutral excitations --  spin
waves\cite{Kallin84,Bychkov81}  which at $\Delta\nu=1$ are described
by the wave functions:
\begin{equation}
\left|SW, {\bf k}\right> =
\sum_{k_1}e^{ik_xk_1\ell^2}\psi^\dagger_{N,k_1,\downarrow}
\psi_{N,k_1-k_y,\uparrow}\left|M_\Phi,\frac{M_\Phi}{2}\right>.
\label{swf}
\end{equation}
Wave functions (\ref{swf}) are  eigenstates of the Hamiltonian
(\ref{He}) with the energies $E_p + E_{SW}(k)$ where the energy of the
spin wave  $E_{SW}(k)$ is given by
\begin{eqnarray}
&\displaystyle{E_{SW}(k) = g_{\rm eff}\hbar\omega_c-}&
 \label{wave}\\
&\displaystyle{\!\int\!\!
\frac{d^2{q}}{(2\pi)^2}\!\frac{2\pi e^2}{\kappa q}
\frac{1}{\varepsilon(q)}
\left[L_N\left(\frac{q^2\ell^2}{2}\right)\right]^2
e^{-q^2\ell^2/2+i{\bf kq}\ell^2}.}&
\nonumber
\end{eqnarray}
For small wave vectors, $kR_c \ll 1$, the spin wave has a quadratic
dispersion relation which is typical for  ferromagnets:
\begin{equation}
E_{SW}(k) = g_0\hbar\omega_c + \frac{e^2R_c}{\pi\kappa}k^2.
\label{wavelong}
\end{equation}
At $k\to 0$ the energy of the spin wave is determined by the bare
$g$-factor in agreement with Larmor's theorem. It is worth mentioning
that the long-wavelength domain of the spin wave spectrum is
controlled by the bare Coulomb interaction. This is because the spatial
scale
$k\ell^2$ important for the spectrum, see Eq.~(\ref{wave}), is much
smaller than the screening radius $a_B$.

In the region $R_c^{-1}\ll k \ll k_F$ the energy of the spin wave reveals
oscillatory behavior
\begin{equation}
E_{SW}(k) = g_{\rm eff}\hbar\omega_c + \frac{e^2}{\pi\kappa
R_c}\left[{\cal C}\left(\frac{k\ell^2}{a_B}\right) -
\frac{\sin(2kR_c)}{2kR_c}\right].
\label{waveint}
\end{equation}
Here ${\cal C}(x)$ is a smooth function with asymptotes
\[
{\cal C}(x) = \left\{
\begin{array}{ll}
\ln x, \quad &x\ll 1\\
-\frac{1}{2x}, \quad &x\gg 1
\end{array} \right..
\]

Spin waves with extremely large wave vectors, $k\gg k_F$, correspond
to almost independent  electron and hole, see Eq.~(\ref{eh}). In this
limit, the energy $E_{SW}$ approaches $\Delta_s$:
\begin{equation}
\Delta_s-E_{SW}(k) =  U(k\ell^2)
\lesssim \frac{\hbar\omega_c}{2N}\ln Nr_s.
\label{veryshortrange}
\end{equation}

\section{Conclusion}
In this paper we constructed a  theory  describing low-energy
excitations in a 2D electron liquid in a  weak magnetic
field. We  have shown that all the excitations with energy smaller than
cyclotron energy $\hbar\omega_c$ can be described by the effective
Hamiltonian acting in the Fock space of the partially filled Landau level
only. Starting from first principles, we obtained the explicit form of
this Hamiltonian by integrating out all the other degrees of freedom.

Armed with this effective theory, we have been able to make
 important predictions:\\
1) We found that the {\em thermodynamic density of states} is negative
for  non-integer filling factors. For a broad range of filling factors,
the value of the thermodynamic density of states was found to be
independent on the value of magnetic field, see
Eq.~(\ref{thermdensity}). This effect may be revealed in
magnetocapacitance measurements\cite{capacitance} in the weak
magnetic field regime.\\
 2) The {\em tunnelling density of states} was
shown to have a gap at the Fermi level. For a broad range of filling
factors, the width of the gap, see Eq.~(\ref{gap2}), was shown to be
consistent with that predicted by hydrodynamic approach of
Ref.~\onlinecite{Aleiner94}. The gap can be observed in tunnelling
experiments\cite{tunnel} in a weak magnetic field.
Evidence of suppression of the tunnelling density of states
for the filling factors $\nu\lesssim 9$ was
 reported recently by Turner {\em et. al.}\cite{weaktunnel}.  This
suppression may be associated with the gap predicted by our theory.
However, the observed width of the gap is twice larger than that
given by Eq.~(\ref{gap2}). \\
3) The {\em exchange enhancement} of the effective $g$-factor  remains
strong in the weak field regime, and at all odd filling factors it takes a
universal value, see Eq.~(\ref{geff}). We found that  the energy scale of
charge excitations is parametrically smaller than the energy scale for
spin excitations. This is qualitatively different from the situation  at
low filling factors, $\nu\lesssim 1$, where both these excitations are
characterized by the same energy scale $e^2/\kappa\ell$.

None of the aforementioned effects could  be obtained in the framework
of the Landau quantization of the spectra of quasiparticles in the
conventional Fermi liquid theory.  This theory describes adequately the
excitations with energies larger than $\hbar\omega_c$.  Thus, our
description of the low-energy properties of interacting 2D electron gas
is complementary to the Fermi liquid picture.

\acknowledgments
Helpful discussions with B.~L.~Altshuler, B.~I.~Halperin, A.~I.~Larkin,
L.~S.~Levitov, Yu.~B.~Lyanda-Geller,  A.~H.~MacDonald, B.~I.~Shklovskii,
and X.~Zhu   are acknowledged with gratitude. We are thankful to
F.~W.~J.~Hekking for critical reading  the manuscript and valuable
comments. This work  was supported by NSF Grant  DMR-9423244.
\appendix
\section{Evaluation of {$\Pi(\omega, q)$}.}
\label{ap1}
Integration in Eq.~(\protect\ref{polarization}) immediately yields
\begin{eqnarray}
&&\displaystyle{\Pi (q,\omega) = -\frac{2m}{\pi}
\sum_{n_1=0}^{N-1}\sum_{n_2=N}^{\infty}
\left[
\frac{(-1)^{(n_2-n_1)}(n_2-n_1)}{(\omega/\omega_c)^2+(n_2-n_1)^2}
\times
\right.
}
\nonumber \nopagebreak\\
&&\displaystyle{\left.
L_{n_1}^{n_2-n_1}\!\left(\frac{{q}^2\ell^2}{2}\right)
L_{n_2}^{n_1-n_2}\!\left(\frac{{q}^2\ell^2}{2}\right)
e^{-{{q}^2\ell^2}/{2}}
\right],
}
\label{pol_arb_k}
\end{eqnarray}
where $L_m^n(x)$ is the Laguerre polynomial\protect\cite{handbook}.

Further simplification is possible for $q \ll k_F$.
Under this condition, we can use the asymptotic expression for the
Laguerre polynomial\protect\cite{handbook}
\begin{equation}
L_n^m(x)\simeq \frac{(n+m)!e^{x/2}{\cal
J}_m\left(\sqrt{2x(2n+m+1)}\right)}
{n!\left[x(n+m/2+1/2)\right]^{m/2}} ,
\label{Laguerre}
\end{equation}
where ${\cal J}_m(x)$ is the Bessel function.
This expression is applicable
if $x \ll n$. Using Eq.~(\ref{Laguerre})
and introducing a new index $j=n_2-n_1$ we approximate
expression (\ref{pol_arb_k}) as:
\begin{eqnarray}
&\displaystyle{\Pi (q,\omega) =
-\frac{2m}{\pi}\sum_{j=1}^{\infty}\left[\frac{j}
{(\omega/\omega_c)^2+j^2} \right.}&\nonumber \\
&\displaystyle{\left.\times\sum^{N-1}_{n_1={\rm max}(N-j,0)}
\left[{\cal J}_{j}
\left(q\ell\sqrt{2n_1+j+1}\right)\right]^2\right].}&
\label{pol1}
\end{eqnarray}
The terms giving the main contribution to the sum in Eq.~(\ref{pol1})
are characterized by value $j\sim
q\ell\sqrt{N}
\ll N$. This enables us to approximate $2n+j+1 \approx 2N$ and perform
the summation over $n_1$. This yields\cite{Simon}
\begin{equation}
\Pi (q,\omega) =
-\frac{2m}{\pi}\sum_{j=1}^{\infty}
\frac{j^2}{(\omega/\omega_c)^2+j^2}\left[{\cal J}_{j}
\left(qR_c\right)\right]^2.
\label{pol2}
\end{equation}
Equation~(\ref{pol1}) can be transformed into an integral form:
substituting the identity
\begin{equation}
\left[{\cal J}_{n}(x)\right]^2=\int_0^\pi \frac{dy}{\pi}\cos n y \
{\cal J}_{0}\left(2x\sin\frac{y}{2}\right)
\label{jsquared}
\end{equation}
into Eq.~(\ref{pol1}) and performing the summation over $j$, one
easily obtains:
\begin{equation}
\Pi (q,\omega)\!=\!
-\frac{m}{\pi}\left[1\! - \!\int_0^\pi
\!\!{dy}{\cal
J}_{0}\!\left(2qR_c\cos\frac{y}{2}\right)
\frac{\frac{\omega}{\omega_c}
\cosh
\frac{\omega}{\omega_c}y}
{\sinh
\pi\frac{\omega}{\omega_c}}\right].
\label{pol3}
\end{equation}
Equation (\ref{pol3}) enables one to obtain the asymptotic form of the
polarization operator in different regimes. In the static limit,
$\omega\ll\omega_c$, one finds from Eqs.~(\ref{pol3})
and (\ref{jsquared}):
\begin{mathletters}
\begin{equation}
\Pi (q,\omega) =-\frac{m}{\pi} \left[1
-{\cal J}_{0}^2\left(qR_c\right) + {\cal
O}\left(\frac{\omega^2}{\omega_c^2}\right)\right].
\label{stat}
\end{equation}
For the high-frequency domain, $\omega \gg \omega_c$, only a small
vicinity of the point $y =\pi$ contributes to
the integral in Eq.~(\ref{pol3}),
and
$\Pi (q,\omega)$ coincides with the result for 2D
electron gas in the absence of the magnetic field:\cite{Stern67}
\begin{equation}
\Pi (q,\omega) =-\frac{m}{\pi} \left[1
-\frac{|\omega|}{\sqrt{\omega^2+q^2v_F^2}} + {\cal
O}\left(\frac{\omega_c^2}{\omega^2}\right)\right].
\label{dynamical}
\end{equation}
In the hydrodynamic limit, $qR_c \ll 1$, one can expand
the Bessel function
in Eq.~(\ref{pol3}) in a Taylor series, which yields
\begin{equation}
\Pi (q,\omega) =-\frac{m}{2\pi}
\left[\frac{q^2v_F^2}{\omega^2+\omega_c^2}+{\cal O}(q^4R_c^4)\right].
\label{hydrodynamic}
\end{equation}
In the opposite limit, $R_c^{-1} \ll q \ll k_F$, the main
contribution to the
integral in Eq.~(\ref{pol3}) comes from the vicinities of points $y=0$
and
$y=\pi$. The calculation gives
\begin{eqnarray}
&\displaystyle{\Pi (q,\omega) \approx -\frac{m}{\pi} \left[1
-B(q,\omega)
\frac{\omega \coth\left
(\pi\frac{\omega}{\omega_c}\right)}{\sqrt{\omega^2+q^2v_F^2}}\
-\right.}&
\nonumber\\
&\displaystyle{\left.
\left(\frac{\omega}{qv_F}\right)\frac{\sin 2qR_c}{\sinh
\pi\frac{\omega}{\omega_c}}\right], }& \nonumber\\
&\displaystyle{B(q,\omega)=1 +\frac{{q^2v_F^2}
\omega_c^2\left(4\omega^2
- q^2 v_F^2\right)}{8\left(\omega^2+q^2v_F^2\right)^3}.
}&
\label{shortrange}
\end{eqnarray}
\end{mathletters}

\section{Parameters of the effective Hamiltonian}
\label{ap0}
In this Appendix we present the details of the derivation of
 the effective
Hamiltonian (\ref{He}).

We start from Eq.~(\ref{almostthere}).
We notice that  operators
$\hat{F}_1$ and $\hat{F}_2^0$, see Eqs.~(\protect\ref{F1}) and
(\protect\ref{F2}), are  integrals of the Matsubara
operators over the time interval $\left[0,\beta_0\right]$. Let us divide
this time interval into $M$ small intervals
$\left[\tau_{m-1},\tau_{m}\right], m=1,2\dots M-1$, where
$\tau_m=m \Delta \tau$ and $\Delta\tau=\beta_0/M$. We imply that
$\Delta\tau$ must be smaller than the characteristic time of the
low-energy dynamics, but much larger than $\omega_c^{-1}$. Then, the
chronologically ordered exponent in Eq.~(\ref{almostthere}) can be
factorized as
\begin{equation}
T_{\tau}\left\{
e^{-\hat{F}_1-\hat{F}_2^0}
\right\}=\prod_{m=1}^M T_{\tau}\left\{e^{-\hat{F}_1^{(m)}-
\hat{F}_2^{(m)}}
\right\},
\label{split}
\end{equation}
where operators $\hat{F}_1^{(m)}$ and $\hat{F}_2^{(m)}$ have the form
\begin{eqnarray}
&&\displaystyle{\hat{F}_1^{(m)}\!=\!\!
\int\!\! d^3\mbox{\boldmath $\xi$}
\phi(\mbox{\boldmath $\xi$})\left(
\hat{\rho}_N (\mbox{\boldmath
$\xi$})-n_e^N\right)
\theta\left([\xi^0-\tau_{m-1}][\tau_{m}-\xi^0]\right),}
\nonumber\\
&&\displaystyle{\hat{F}_2^{(m)}
=\!\int\! d^3\mbox{\boldmath
$\xi_1$}
\int\! d^3\mbox{\boldmath $\xi_2$}
\left[
\bar{\Psi}_N(\mbox{\boldmath
$\xi_1^+$})\Psi_N(\mbox{\boldmath
$\xi_2$}) - 2G_N(\mbox{\boldmath $\xi_2$},
\mbox{\boldmath $\xi_1$})
\right] \times} \nonumber \\
&&\displaystyle
{\quad \phi(\mbox{\boldmath $\xi_1$})
\tilde{G}_{0}(\mbox{\boldmath $\xi_1$},\mbox{\boldmath $\xi_2$})
\phi(\mbox{\boldmath
$\xi_2$})
\theta\left([\xi^0_1-\tau_{m-1}][\tau_{m}-\xi^0_1]\right).
}
 \label{F2A}
\end{eqnarray}
Definitions (\ref{F2A}) differ from the corresponding  definitions
(\protect\ref{F1}) and (\protect\ref{F2}) by the constraint on the
time domain of integration. We do not need to impose an additional
constraint on the  time component of
\mbox{\boldmath
$\xi_2$} in the second of Eqs.~(\ref{F2A})
because
$\tilde{G}_0$ decays rapidly at
$|\xi_1^0-\xi_2^0| > 1/\omega_c$, see Eq.~(\ref{G0}).

Now, we substitute Eq.~(\ref{split}) into
Eq.~(\protect\ref{almostthere})
and perform the integration over the field $\phi$. It yields
\begin{equation}
\hat{\Lambda}(\beta_0)=
e^{-\beta_0(\Omega_0 +\Delta\Omega +\hat{H}^N_0)}
\prod_{m=1}^M \left<T_{\tau}
\left\{e^{-\hat{F}_1^{(m)}-\hat{F}_2^{(m)}}
\right\}\right>,
\label{LambdaAp}
\end{equation}
where average $\left<{\cal F}\right>$ of an arbitrary functional
${\cal
F}\left\{\phi\right\}$  over the field $\phi$ is
defined as
\begin{eqnarray}
&\displaystyle{\left<{\cal F}\right>=
e^{\beta_0\Delta\Omega}\!\int\!{\cal
D}\phi\ {\cal F}\left\{\phi\right\}\times}& \nonumber \\
&\displaystyle{\exp\left(\frac{1}{2}\!\int\! d^3\mbox{\boldmath
$\xi_1$}\!\int\!
 d^3\mbox{\boldmath $\xi_2$}
\phi(\mbox{\boldmath $\xi_1$})
\phi(\mbox{\boldmath $\xi_2$})
{\cal S}(\mbox{\boldmath $\xi_1\! -\!\xi_2$})\right).}&
\label{average}
\end{eqnarray}
The normalization factor  $e^{\beta_0\Delta\Omega}$ describes
the fluctuations of field $\phi$ around
the saddle point:
\begin{equation}
e^{-\beta_0\Delta\Omega}\!\!=
\!\int\!\!{\cal
D}\phi
\exp\!\left(\frac{1}{2}\!\int\!\!d^3\!\mbox{\boldmath $\xi$}\!\!\int\!\!
d^3\!\mbox{\boldmath $\xi^\prime$}
\phi(\mbox{\boldmath $\xi$})
\phi(\mbox{\boldmath $\xi^\prime$})
{\cal S}(\mbox{\boldmath $\xi\! -\!\xi^\prime$})\right).
\label{DeltaOmega}
\end{equation}
An explicit expression for  $\Delta\Omega$ is presented in
Subsection \ref{calomega}.

Because $\Delta\tau$ is small, we can expand every factor in
Eq.~(\ref{LambdaAp}) up to the first order in $\Delta\tau$:
\begin{equation}
\left<T_{\tau}\left\{e^{-\hat{F}_1^{(n)}-\hat{F}_2^{(n)}}\right\}\right>
\approx 1 - \Delta\tau \hat{ H}_I
\left\{\bar{\Psi}_N(\tau_n), \Psi_N(\tau_n)\right\}.
\label{expand}
\end{equation}
This relation defines  the operator  $\hat{H}_I$.
Let us notice that this operator actually does not depend on  $\tau_n$
because it contains the same number of creation and annihilation
operators taken at the same moment of time.
Having this in mind, we substitute
Eq.~(\ref{expand}) into Eq.~(\ref{LambdaAp}) and obtain
\begin{equation}
\hat{\Lambda}(\beta_0)=e^{-\beta_0(\Omega_0 +\Delta\Omega +
\hat{H}^N_0)}
\left(1 - \frac{\beta_0}{M}\hat{H}_I\right)^M.
\label{LambdaAp1}
\end{equation}
Finally, we take the  limit $M\to\infty$ in Eq.~(\ref{LambdaAp1}) and
substitute the result in Eq.~(\protect\ref{Heff}). It yields
\begin{equation}
\hat{H}_{eff} = \Omega_0 + \Delta\Omega + \hat{H}_0^N + \hat{ H}_I.
\label{apanswer}
\end{equation}

Now, one has to perform the actual calculation of the average in the
left hand side of Eq.~(\ref{expand}) in order to find the operator
$\hat{H}_I$. It can be done by using perturbation theory.  The leading
terms in the small parameters
$r_s$ and $1/N$ are
\begin{eqnarray}
\left<T_{\tau}\left\{e^{-\hat{F}_1^{(n)}-\hat{F}_2^{(n)}}\right\}\right>
\approx
1 - \left<\hat{F}_1^{(n)}+\hat{F}_2^{(n)}\right> \nonumber\\
+ \frac{1}{2}
\left<T_{\tau}\left(\hat{F}_1^{(n)}+\hat{F}_2^{(n)}\right)^2\right>.
\label{perturbation}
\end{eqnarray}
To find the operator $\hat{ H}_I$, only terms  linear in $\Delta \tau$
should be retained.

Calculation of the right-hand  side of Eq.~(\ref{perturbation}) is
 carried out
with the help of Eqs.~(\ref{F2A})
and of the correlation functions of the field $\phi$:
\begin{eqnarray}
&\displaystyle{\left<\phi\right>=0,}& \label{corphi}\\
&\displaystyle{
\left<\phi (\mbox{\boldmath $r$},\tau)\phi(0,0)\right>=
-\frac{e^2}{\kappa |r|}\delta (\tau) -}& \nonumber \\
&\displaystyle{\int \frac{d\omega d^2{q}}
{(2\pi)^3}\left(\frac{2\pi e^2}{\kappa q}\right)^2
\frac{\Pi(q,\omega)}{\varepsilon(q,\omega)}
e^{i{\bf qr}-i\omega\tau}.}&
\nonumber
\end{eqnarray}
The latter expression follows from Eq.~(\ref{average}) and
(\ref{renorm}).
In the second of Eqs.~(\ref{corphi}), we explicitly separated a
term which is singular at $\tau = 0$.

The  averages  that give the leading contribution
to the effective Hamiltonian   are
\begin{eqnarray}
&&\displaystyle
{\left<\hat{F}_2^{(n)}\right> = \Delta\tau\left(\mu_{ex}+\mu_c\right)
\int d^2\mbox{\boldmath $r$}
\hat{\rho}_N(\mbox{\boldmath $r$}),}
\label{F2av} \\
&&\displaystyle{
\frac{1}{2}\left<T_{\tau}\left(\hat{F}_1^{(n)}\right)^2\right>=
- \Delta\tau \left(
\hat{H}_{int}^{eff}
+\mu_{p}
\int d^2\mbox{\boldmath $r$}
\hat{\rho}_N(\mbox{\boldmath $r$})
\right),}
\nonumber
\end{eqnarray}
where the operator $\hat{H}_{int}^{eff}$ is defined by Eq.~(\ref{heint}).

In Eqs.~(\ref{F2av}), $\mu_{ex}$ coincides with the
well-known exchange correction to the chemical potential:
\begin{equation}
\mu_{ex} = -2 \pi\ell^2 \int d^2\mbox{\boldmath $r$}
\frac{e^2}{\kappa r}
\tilde{G}_0\left(\mbox{\boldmath $r$}, 0; 0, \epsilon\to +0\right)
P_N\left(0, \mbox{\boldmath $r$}\right),
\label{muex}
\end{equation}
where $\tilde{G}_0$ is defined by Eq.~(\ref{G0}),
and $\mu_{c}, \mu_p$ are  the correlation
corrections to the chemical potential found in the random
phase approximation:
\begin{eqnarray}
&&\displaystyle{\mu_c=- \int\!\frac{d\omega d^2{q}}
{(2\pi)^3}\!\left(\frac{2\pi e^2}{\kappa q}\right)^2\!
\frac{\Pi(q,\omega)}{\varepsilon(q,\omega)}
\sum_{n \neq N}\!
\frac{\omega_c(N-n)}{\omega^2+\omega_c^2(N-n)^2}}
\nonumber\\
&&\displaystyle{\times(-1)^{(N-n)}
L_{n}^{N-n}\!\left(\frac{{q}^2\ell^2}{2}\right)
L_{N}^{n-N}\!\left(\frac{{q}^2\ell^2}{2}\right) e^{-{{q}^2\ell^2}/{2}},
} \label{muc}\\
&&\displaystyle{
\mu_p = \pi \ell^2
\int d^2\mbox{\boldmath $r$} P_N\left(0, \mbox{\boldmath $r$}\right)
P_N\left( \mbox{\boldmath $r$}, 0\right)
\left(U(r) - \frac{e^2}{\kappa r}\right).}
\label{mup}
\end{eqnarray}
When deriving Eqs.~(\ref{muex}),  (\ref{muc}), we use
 Eqs.~(\ref{equal_times}) and (\ref{G0}).

Finally, substitution of  Eq.~(\ref{F2av}) into Eq.~(\ref{perturbation})
enables us to find the operator $\hat{H}_I$. Comparing
Eq.~(\ref{apanswer}) and Eq.~(\ref{He}), we obtain  the expressions for
the parameters of  the Hamiltonian
$H_{eff}$:
\begin{eqnarray}
&\Omega = \Omega_0 + \Delta\Omega,&
\label{omegai}\\
&\mu^\ast = \mu - N \hbar\omega_c - \mu_{ex} - \mu_c - \mu_p.&
\label{mustar}
\end{eqnarray}

The following subsections are devoted to the explicit calculation of the
parameters of the effective Hamiltonian in terms of the filling factor
and interaction strength. We will assume that  condition
(\protect\ref{cond3}) is met.

\subsection{Calculation of $\Delta\Omega$.}
\label{calomega}
Equations~(\ref{DeltaOmega}), (\ref{normalization})
and (\ref{renorm})
yield
\begin{equation}
\frac{\Delta\Omega}{L_xL_y} = \frac{1}{2}
 \int \frac{d\omega d^2{q}}
{(2\pi)^3}\ln\left({\varepsilon(q,\omega)}\right).
\label{DO}
\end{equation}
With the help of Eqs.~(\ref{fullepsilon}), (\ref{polarization}) and
(\ref{equal_times}), one can transform Eq.~(\ref{DO}) into a more
convenient form:
\begin{eqnarray}
&&\Delta\Omega = \Delta\Omega_{ex} + \Delta\Omega_c,
 \label{parts} \\
&&\displaystyle{\frac{\Delta\Omega_{ex}}{L_xL_y}=-\int
d^2\mbox{\boldmath $r$}
\frac{e^2}{\kappa r}
\tilde{G}_0\left(\mbox{\boldmath $r$}, 0; 0,
\epsilon\right)\tilde{G}_0\left(0, 0; \mbox{\boldmath $r$},
\epsilon\right),}
\label{Omex}\\
&&\displaystyle{\frac{\Delta\Omega_{c}}{L_xL_y}=-\frac{1}{2}\!
\int_0^1\! d\alpha \!\int\! \frac{d\omega d^2{q}}
{(2\pi)^3}\left(\frac{2\pi e^2}{\kappa q}\right)^2
\frac{\alpha\Pi^2(q,\omega)}{1-\alpha\frac{2\pi e^2}{\kappa
q}\Pi(q,\omega)}. } \nonumber\\
\label{Omc}
\end{eqnarray}
Expression (\ref{Omex}) corresponds to the first order exchange
correction to the ground state energy, and $\epsilon \to +0$. Equation
{}~(\ref{Omc}) is the  correlation correction to the ground state energy
equivalent to the sum of  ring diagrams\cite{review}.

First, we evaluate the exchange energy (\ref{Omex}). Using the explicit
 expression for  the Green's function (\ref{G0}), the property of the
Landau level wave functions,
\begin{eqnarray}
&&P_n({\bf
r}_1,{\bf
r}_2)=\sum_{k}\varphi_{n,k}^{\ast}({\bf r}_2)\varphi_{n,k}({\bf
r}_1)=\label{property}\\
&&\frac{1}{2\pi\ell^2}e^{i(y_1-y_2)(x_1+x_2)/2\ell^2-|{\bf r}_2-{\bf
r}_1|^2/4\ell^2}L_n\left(\frac{|{\bf r}_2-{\bf r}_1|^2}{2\ell^2}\right),
\nonumber
\end{eqnarray}
and the identity for the  Laguerre polynomials,
\begin{equation}
\sum_{i=0}^nL_i(x) = L_n^1(x),
\label{lid}
\end{equation}
 we obtain
\begin{equation}
\frac{\Delta\Omega_{ex}}{L_xL_y}=-\frac{e^2}{\kappa
\ell\left(2\pi\ell^2\right)}
\int_0^\infty dx
e^{-x^2/2}\left[L_{N-1}^1\left(\frac{x^2}{2}\right)\right]^2.
\label{omex1}
\end{equation}
Integration in Eq.~(\ref{omex1}) can be easily performed,
which yields for $N \gg 1$:
\begin{equation}
\frac{\Delta\Omega_{ex}}{L_xL_y}\!=-\frac{e^2}{\kappa
\pi^2\ell^3}\!
\left[\frac{2}{3}\left(2N\right)^{3/2}\!+\! \frac{\ln N}{16(2N)^{1/2}} \!
+\!
{\cal O}\left(\frac{1}{N^{1/2}}\right)\right].
\label{omex2}
\end{equation}

The first term in brackets coincides with the exchange energy of 2D
electron gas at zero magnetic field, and the second term appears due to
the confinement of the electron wave functions by the magnetic field.
The logarithmical factor $\ln N \sim \ln\left(k_FR_c\right)$  arises
due to the integration over   relatively large spatial scales
$k_F^{-1} \ll r \lesssim R_c$  in Eq.~(\ref{Omex}).  On these scales, the
screened interaction potential is significantly smaller than the bare
potential appearing in Eq.~(\ref{Omex}). Therefore, it is plausible to
anticipate that the correlation term of the thermodynamic potential,
which accounts for the screening effect, should partially compensate
the large logarithmic factor in Eq.~(\ref{omex2}), and lead to the
replacement
$R_c\to a_B$ in the argument of the logarithm. The calculation of the
correlation energy (\ref{Omc}) explicitly demonstrates this.

In zero magnetic field the  contribution  of the correlation
energy\cite{review}  is only of the order of ${\cal O}( r_s^2)$.
The situation changes in the magnetic field, where a contribution
proportional to $r_s\ln (Nr_s)$ appears. This term arises mainly from
integration over the domain of wave vectors
$R_c^{-1} \ll q \ll a_B^{-1}$  in Eq.~(\ref{Omc}), where we can use the
asymptotic expression (\ref{shortrange}) for $\Pi(q,\omega)$. It yields
with the logarithmic accuracy
\begin{equation}
\frac{\Delta\Omega_{c}}{L_xL_y}=\frac{e^2}{\kappa
\ell\sqrt{2N}\left(16\pi\ell^2\right)}
\ln\left( N r_s\right) + {\cal O}(r_s^2).
\label{omec2}
\end{equation}

Finally, with the help of Eqs.~(\ref{omex2}) and (\ref{omec2}), we
obtain for the thermodynamic potential
(\ref{omegai}),
\begin{eqnarray}
&\displaystyle{{\Omega}=\frac{L_xL_y}{\pi\ell^2}\left\{
\hbar\omega_c\frac{N(N-1)}{2} - \mu N
- \right.}& \nonumber\\
&\displaystyle{
\left.
\frac{e^2}{\pi\kappa\ell}\left[\frac{2}{3}(2N)^{3/2} +
\frac{1}{16 (2N)^{1/2}}\ln\left(r_s^{-1}\right)
\right]
\right\}.}&
\label{Omegafinal}
\end{eqnarray}
This result contradicts Ref.\onlinecite{Isihara79} where the
contributions (\ref{Omex})  and (\ref{Omc})  were evaluated and a
correction $\propto r_s^{3/4}$ was obtained instead of $r_s |\ln r_s|$.

\subsection{Calculation of $\mu^\ast$.}
\label{calmu}
We start from the calculation of the exchange correction to the
chemical potential. With the help of Eqs.~(\ref{G0}), (\ref{property})
and (\ref{lid}) we transform Eq.~(\ref{muex}) to the form
\begin{equation}
\mu_{ex}=-\frac{e^2}{\kappa \ell}
\int_0^\infty dx
e^{-x^2/2}L_{N-1}^1\left(\frac{x^2}{2}\right)
L_{N}\left(\frac{x^2}{2}\right).
\label{muex1}
\end{equation}
After integration in Eq.~(\ref{muex1}) we obtain for $N
\gg 1$ with logarithmic accuracy:
\begin{equation}
\mu_{ex}=-\frac{e^2}{\kappa \ell}
\left(\frac{2}{\pi}\left(2N\right)^{1/2} - \frac{1}{2\pi(2N)^{1/2}}
\ln N
\right).
\label{muex2}
\end{equation}
The correlation shift of the chemical potential, Eq.~(\ref{mup}), is
calculated with the help of Eqs.~(\ref{property}) and
(\ref{renpotential}). With the logarithmic accuracy, the calculation
gives
\begin{equation}
\mu_{p}=-\frac{e^2}{\kappa \ell}
 \frac{1}{2\pi(2N)^{1/2}}\ln \left( Nr_s\right).
\label{mup2}
\end{equation}
It can be checked by an explicit calculation that the other correlation
correction to the chemical potential,
$\mu_c$, see Eq.~(\ref{muc}), contains an additional small factor in
comparison with Eq.~(\ref{mup2}): $\mu_c \simeq \mu_p/N$, and,
therefore, it can be  neglected. Finally,  with the logarithmic accuracy,
we have for $\mu^\ast$ appearing in Eq.~(\ref{mustar})  the result:
\begin{equation}
\mu^{\ast}=\mu-N\hbar\omega_c+\frac{e^2}{\kappa \ell}
\left(\frac{2}{\pi}\left(2N\right)^{1/2} - \frac{\ln\left(r_s^{-1}\right)
}{2\pi(2N)^{1/2}}
\right).
\label{Mufinal}
\end{equation}

\narrowtext

\mediumtext

\begin{references}
\bibitem{gfactor} Previously, the effect of screening on the effective
$g$ - factor was considered in Refs.~\onlinecite{MacDonald92} and
\onlinecite{Ando74}, however, no explicit analytical expression was
given.
\end{references}
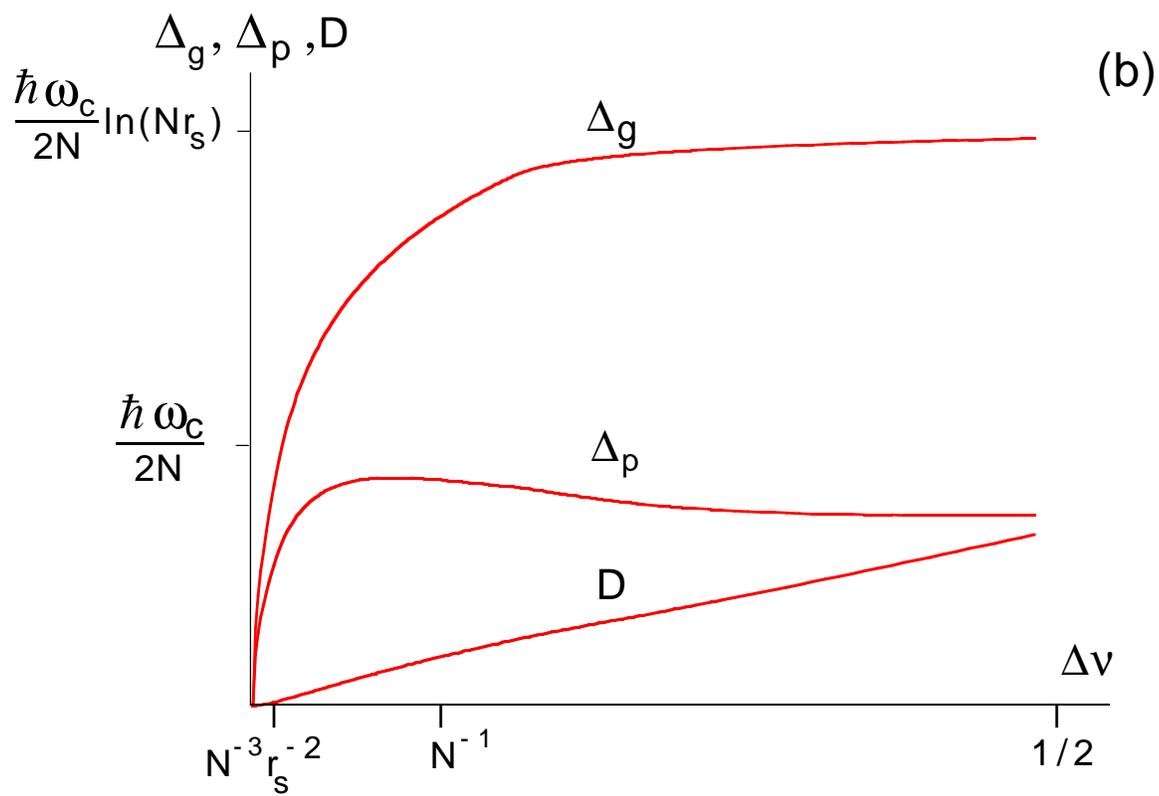
\end{document}